\documentclass[aps,pre,showpacs,twocolumn,groupedaddress]{revtex4}
\usepackage{latexsym}
\usepackage{graphicx}
\usepackage{amsmath}
\usepackage{amssymb}
\usepackage{bm}
\begin{document}
\title{Ornstein-Zernike equation and Percus-Yevick theory for
	 molecular crystals}
\author{Michael Ricker}
\email{mricker@uni-mainz.de}
\author{Rolf Schilling}
\email{rschill@uni-mainz.de}
\affiliation{Institut f\"ur Physik, Johannes Gutenberg-Universit\"at
Mainz, Staudinger Weg 7, D-55099 Mainz, Germany}
\date{\today}
\begin{abstract}
We derive the Ornstein-Zernike equation
for molecular crystals of axially symmetric particles and
apply the Percus-Yevick approximation to this system. 
The one-particle orientational distribution
function $\rho^{(1)}(\Omega)$ has a nontrivial dependence on the
orientation $\Omega$, in contrast to a liquid, and is needed as an
input. Despite some differences, the Ornstein-Zernike equation for
molecular crystals has a similar structure as for liquids. We solve
both equations numerically for hard ellipsoids of revolution on a simple cubic
lattice. Compared to molecular liquids, the orientational
correlators in direct and reciprocal space exhibit less
structure. However, depending on the lengths $a$ and $b$ 
of the rotation axis and the perpendicular 
axes of the ellipsoids, respectively, different behavior
is found. For oblate and prolate ellipsoids with
$b \gtrsim 0.35$ (in units of the lattice constant),
damped oscillations in distinct directions of direct
space occur for some of the orientational correlators.
They manifest themselves in some of the correlators in
reciprocal space as a maximum at the Brillouin zone edge,
accompanied by a maximum at the zone center for other correlators. 
The oscillations indicate alternating orientational fluctuations, 
while the maxima at the zone
center originate from nematic-like orientational fluctuations. For 
$a \lesssim 2.5$ and $b \lesssim 0.35$, the oscillations
are weaker, leading to no marked maxima at the 
Brillouin zone edge. For $a \gtrsim 3.0$ and $b \lesssim 0.35$, 
no oscillations occur any longer. For many of the orientational correlators in
reciprocal space, an increase
of $a$ at fixed $b$ or vice versa leads to a divergence at the zone center
${\bf q}={\bf 0}$, consistent with the formation of nematic-like long range
fluctuations, and for some oblate and prolate systems with $b
\lesssim 1.0$ a simultaneous tendency to divergence of few other
correlators at the zone edge is observed. Comparison of the
orientational correlators with those from
MC simulations shows satisfactory agreement. From these simulations
we also obtain a phase boundary in the $a-b$-plane for order-disorder
transitions. 
\end{abstract}
\maketitle
\section{\label{secI}Introduction}
The experimental, numerical and analytical study of structural
properties of simple liquids is a well established discipline of
condensed matter physics. In the center of such investigations is the
static structure factor $S(q)$. There are powerful integral equations
allowing an approximate calculation of $S(q)$~\cite{HMD1990}. The
starting point is the Ornstein-Zernike (OZ) equation, relating the
total correlation function $h(q)$ and the direct correlation function
$c(q)$. An additional closure relation, like the Percus-Yevick (PY)
approximation, then allows to determine $h(q)$, from which $S(q)$
follows from 
\begin{equation}
\label{eq1}
S(q) = 1 + \rho h(q),
\end{equation}
where $\rho$ is the number density of the liquid. Application of the
PY approximation to a liquid of hard spheres yields good agreement
with the exact result for intermediate values of
$\rho$~\cite{HMD1990}. However, the crystallization of hard spheres cannot be
described by PY theory.

The extension of the OZ equation and the PY approximation (or other
closure relations) to molecular liquids is
straightforward~\cite{HMD1990,GG1984} and has been applied
extensively (see, e.g.,~\cite{RSS1994,LL1999}). As for simple liquids, PY theory
usually does not yield an order-disorder phase transition. Therefore, 
it was quite surprising that a recent application of the molecular
version of that theory to a liquid of hard ellipsoids of revolution
with aspect ratio $X_{0}$ has allowed the location of a phase boundary in
the $\rho-X_{0}$-plane, at which a transition to a nematic phase takes
place~\cite{LL1999}. 

Much less analytical work exists for molecular
crystals~\cite{JDW2001,SHER1979}. These are crystalline materials
with, e.g., a molecule at each lattice site. One of the main interests
concerns phase transitions of the translational and rotational degrees
of freedom (see, e.g., the review~\cite{MUE1998}). These transitions
are influenced by the translation-rotation coupling~\cite{LM1994}. But
one also expects phase transitions if the crystal is assumed to be
rigid. For instance, let us fix hard ellipsoids of revolution with
their centers at the lattice sites of, e.g., a simple cubic lattice
with lattice constant equal to one. With $a$ and $b$ the lengths 
of the rotation axis and the perpendicular axes of the ellipsoids, 
respectively, an increase of $a$ for fixed $b$ may yield at a critical
length $a_{c}(b)$ a transition to an orientationally ordered phase,
because the entropy of the ordered phase is larger than that of the
disordered phase. 

Besides phase transitions, the study of the orientational structure of
molecules on a rigid lattice is of interest, too. Changing temperature
will influence the steric hindrance between the molecules. The same
happens for hard ellipsoids when changing $a$ and $b$. Analogous to
simple and molecular liquids, one can quantify such static
orientational properties by the one-particle orientational
distribution
$\rho^{(1)}(\Omega)$~\cite{SHER1979,YP1980,LM1994,BP1988,BP1989,BP1994} and
by the orientational correlation function $G_{nn'}(\Omega,\Omega')$,
where the orientation $\Omega$ can be characterized, e.g., by the
Euler angles $(\phi,\theta,\chi)$ or, for axially symmetric particles,
by $(\phi,\theta)$. Then the following questions arise: How to compute 
$\rho^{(1)}(\Omega)$ and, above all, the correlation function
$G_{nn'}(\Omega,\Omega')$ by an analytical method? Does the result
for $G_{nn'}(\Omega,\Omega')$ allow to locate a phase boundary where a
transition to an orientationally ordered phase occurs? To provide
answers to these questions is the main motivation of our contribution.

Our paper is organized as follows. In Sec.~\ref{secII} we will
introduce the model and the basic physical quantities like the
one-particle distribution function $\rho^{(1)}(\Omega)$ and the
orientational correlation function $G_{nn'}(\Omega,\Omega')$. The
analytical approach of calculating $G_{nn'}(\Omega,\Omega')$ or its
transform $S_{\lambda\lambda'}({\bf q})$ from the OZ equation in
combination with the PY approximation is described in
Sec.~\ref{secIII}. Results from PY theory for hard ellipsoids of
revolution on a simple cubic lattice will be shown in Sec.~\ref{secIV}
and compared with those from MC simulations. This section also
presents a short discussion of the phase transition for the ellipsoids
from an orientationally disordered to an ordered phase. The final
Sec.~\ref{secV} contains a discussion of the results and some
conclusions. We add a couple of appendices, including extensive technical
manipulations needed in Secs.~\ref{secIII},~\ref{secIV}.
\section{\label{secII}Distribution and correlation functions and
their symmetries}
\subsection{\label{secIIa}One- and two-particle distribution and correlation functions}
We consider a three-dimensional periodic lattice with $N$ lattice
sites and periodic boundary conditions. If the $n$-th lattice site has
the position ${\bf x}_{n}$, the difference between two sites is
the vector ${\bf x}_{nn'} = {\bf x}_{n'} - {\bf x}_{n}$. We assume the
lattice to be rigid with lattice constant equal to one. At each
lattice site we fix a rigid molecule, not necessarily with its center
of mass. Restricting to linear molecules, the orientation of the
molecule at site $n$ is given by $\Omega_{n} = (\phi_{n},\theta_{n})
\in S^2$. The third Euler angle $\chi_{n}$ is irrelevant for our
purposes. Extension of our theoretical approach below to arbitrary
molecules is straightforward. The interaction energy
$V(\{\Omega_{n}\})$ is assumed to be pairwise and the classical
Hamiltonian is given by
\begin{equation}
\label{eq2}
H(\{\Omega_{n} \}, \{ {\bf l}_{n} \}) = \sum\limits_{n =1}^{N}
\frac{1}{2} {\bf l}_{n}^{T} {\bf I}^{-1}(\Omega_{n}) {\bf l}_{n} +
V(\{\Omega_{n} \})\,,
\end{equation}
where ${\bf l}_{n}$ and ${\bf I}(\Omega_{n})$, respectively, are the
angular momentum and the tensor of inertia of the molecule at site
$n$ in the space fixed frame. Since we will investigate
static quantities only, the kinetic part of $H(\{\Omega_{n} \}, \{
{\bf l}_{n} \})$ does not matter. 

In order to describe the orientational degrees of freedom, we
introduce the microscopic one particle density
$\rho_{n}(\Omega)$, $\Omega \in S^2$, at lattice site
$n$ and its associated fluctuation defined by
\begin{subequations}
\label{eq3}
\begin{align}
\label{eq3a}
\rho_{n}(\Omega) &= \delta(\Omega|\Omega_{n})\,,\\
\label{eq3b}
\delta \rho_{n}(\Omega) &= \rho_{n}(\Omega) -
\langle \rho_{n}(\Omega)\rangle \,,
\end{align}
\end{subequations}
where $\delta(\Omega|\Omega') = \sin \theta \,\delta (\theta -\theta')
\delta (\phi -\phi')$, and $\langle ( \cdot ) \rangle$ denotes
canonical averaging with respect to $H(\{\Omega_{n} \}, \{
{\bf l}_{n} \})$. Note that the $\Omega_{n}$-dependence of $\rho_{n}$
and $\delta \rho_{n}$ is suppressed. The one-particle orientational
distribution function is given by
\begin{equation}
\label{eq4}
\rho^{(1)}(\Omega) = \langle \rho_{n}(\Omega) \rangle \,,
\end{equation}
which is $n$-independent due to the lattice translational invariance
of $H(\{\Omega_{n} \}, \{ {\bf l}_{n} \})$, and the two-particle
distribution is defined as
\begin{equation}
\label{eq5} 
\rho^{(2)}_{nn'}(\Omega,\Omega') \equiv
\rho^{(2)}_{{\bf x}_{nn'}}(\Omega,\Omega') = \langle \rho_{n}(\Omega)
\rho_{n'}(\Omega')  \rangle \,, \quad n \neq n'\,.
\end{equation}
Note that $\rho^{(2)}_{nn'}$ is defined for $n \neq n'$, only, and that 
the index pair $nn'$ is fully equivalent to the vector ${\bf x}_{nn'}$.
Making use of Eqs.~(\ref{eq3}), it follows
\begin{equation}
\label{eq6}
\int\limits_{S^2} \rho^{(1)}(\Omega) \, d\Omega =  1\,,
\end{equation}
\begin{subequations}
\label{eq7}
\begin{align}
\label{eq7a}
\int\limits_{S^2} \rho^{(2)}_{nn'}(\Omega,\Omega') \, d\Omega   & =
\rho^{(1)}(\Omega')&\quad (n \neq n')\,,\\
\label{eq7b}
\int\limits_{S^2} \rho^{(2)}_{nn'}(\Omega,\Omega') \, d\Omega'   & =
\rho^{(1)}(\Omega)&\quad (n \neq n') \,.
\end{align}
\end{subequations}

\begin{figure*}[t]
\includegraphics[angle=270,width=16.5cm]{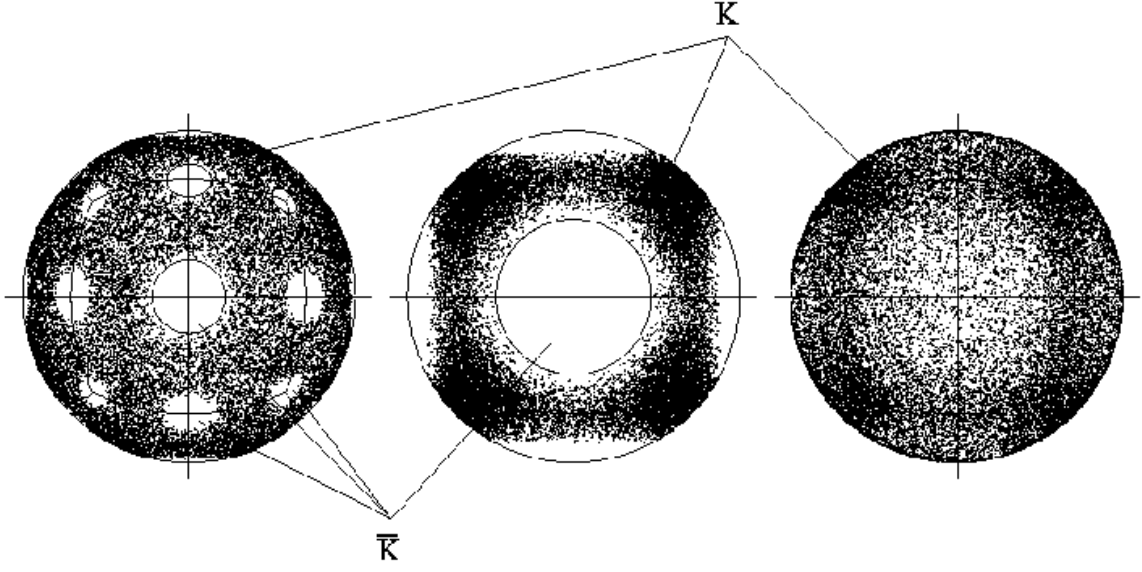}
\caption{\label{fig1} MC results for $\rho^{(1)}(\Omega)$ for hard ellipsoids of revolution 
with $a = 3.6$, $b = 0.24$ (left), $a = 1.2$, $b = 0.88$ (middle) and 
$a = 0.56$, $b = 1.1$ (right) on a sc lattice. Orientations
on $S^{2}$ obtained from MC runs are projected along the fourfold lattice direction. 
Parts of $\bar{K}$ occur along the two-, three- and fourfold lattice
directions, depending approximately on whether $\frac{a+b}{2}$ exceeds the site-site 
spacing along one of these directions. Some circles approximating the edges of the 
parts of $\bar{K}$ are shown as an aid for the eye.}
\end{figure*}

It is important to realize that in contrast to isotropic molecular
liquids the one-particle distribution $\rho^{(1)}(\Omega)$ depends 
on $\Omega$. An additional important feature may occur for
particles with hard core interaction, e.g. hard ellipsoids, which are
big enough. In that case there may exist a non-empty subset $\bar{K}$
of $S^2$ such that $\rho^{(1)}(\Omega) = 0$ for all $\Omega \in
\bar{K}$. Hence, the orientation of each particle is restricted to the
complement $K = S^2 \setminus \bar{K} \subset S^2$. In the following, we
need to distinguish between the two cases $K = S^2$ and $K \subset S^2$,
where $K$ is the area on $S^2$ which can be reached by particle
orientations. Clearly, it is
$\rho^{(2)}_{nn'}(\Omega,\Omega')=0$, if  $\rho^{(1)}(\Omega) = 0$ or  
$\rho^{(1)}(\Omega') = 0$, but not vice versa.
Three examples for $\rho^{(1)}(\Omega)$
are shown in Fig.~\ref{fig1}, obtained from MC simulations for hard 
ellipsoids on a simple cubic lattice.                                 

Next, we introduce the orientational density-density correlation
function $G_{nn'}(\Omega,\Omega')$. It is the correlation of the
fluctuations of $\rho_{n}(\Omega)$ at lattice sites $n$ and $n'$: 
\begin{equation}
\label{eq8}
G_{nn'}(\Omega,\Omega') = \langle \delta\rho_{n}(\Omega)
\delta\rho_{n'}(\Omega') \rangle \, .
\end{equation}
By use of Eqs.~(\ref{eq3})-(\ref{eq5}) we get 
\begin{align}
\label{eq9}
G_{nn'}(\Omega,\Omega') = & \; \delta_{nn'} \rho^{(1)}(\Omega)
\,\delta(\Omega|\Omega') -
\rho^{(1)}(\Omega) \rho^{(1)}(\Omega') \nonumber \\
& + (1-\delta_{nn'}) \rho^{(2)}_{nn'}(\Omega,\Omega')  \,.
\end{align}
$G_{nn'}(\Omega,\Omega') = G^{(s)}_{nn'}(\Omega,\Omega') + 
G^{(d)}_{nn'}(\Omega,\Omega')$ consists of a self part and a distinct
part, which are explicitely
\begin{subequations}
\label{eq10}
\begin{align}
\label{eq10a}
G^{(s)}_{nn'}(\Omega,\Omega') & = \delta_{nn'} \left(
\rho^{(1)}(\Omega) \delta(\Omega|\Omega') -
\rho^{(1)}(\Omega) \rho^{(1)}(\Omega') \right) \,, \\
\label{eq10b}
G^{(d)}_{nn'}(\Omega,\Omega') &= \left( 1 - \delta_{nn'} \right)
\left( \rho^{(2)}_{nn'}(\Omega,\Omega') -
\rho^{(1)}(\Omega) \rho^{(1)}(\Omega')\right) \,.
\end{align}
\end{subequations}
Due to the properties~(\ref{eq6}) and~(\ref{eq7}) of the particle
distribution functions, $G_{nn'}(\Omega,\Omega')$ fulfills for all $nn'$
\begin{equation}
\label{eq11}
\int\limits_{S^2} G_{nn'}(\Omega,\Omega') \, d\Omega =
\int\limits_{S^2} G_{nn'}(\Omega,\Omega') \, d\Omega' = 0\,. 
\end{equation}

For the lattice system, the pair and total correlation functions
$g_{nn'}(\Omega,\Omega')$ and $h_{nn'}(\Omega,\Omega')$ are
introduced in the same manner as for a liquid~\cite{HMD1990,GG1984}: 
\begin{align}
\label{eq12}
g_{nn'}(\Omega,\Omega') & = \frac{\rho^{(2)}_{nn'}(\Omega,\Omega')}
{\rho^{(1)}(\Omega)\, \rho^{(1)}(\Omega')}& \quad (n \neq n') \,, \\
\label{eq13}
h_{nn'}(\Omega,\Omega') & = g_{nn'}(\Omega,\Omega') -1 &\quad (n \neq
n') \,.
\end{align}
Here, some comments are in order. First, $g_{nn'}$ and $h_{nn'}$ are 
defined for $n \neq n'$ only, and if $K \subset S^2$, 
for $\Omega,\Omega' \in K$ only. Second, $h_{nn'}(\Omega,\Omega')$ in
general obeys 
\begin{equation}
\label{eq14}
\int\limits_{S^2} h_{nn'}(\Omega,\Omega') \,d\Omega \neq 0 \,, \quad  
\int\limits_{S^2} h_{nn'}(\Omega,\Omega') \, d\Omega' \neq 0 
\quad (n \neq n') \,.
\end{equation}
The same is true for $g_{nn'}(\Omega,\Omega')$. Third, due to 
Eqs.~(\ref{eq6}),~(\ref{eq7}),~(\ref{eq12}) and~(\ref{eq13}) it is
\begin{subequations}
\label{eq15}
\begin{align}
\int\limits_{S^2} \rho^{(1)}(\Omega) \,h_{nn'}(\Omega,\Omega') 
\,d\Omega & = 0 \quad (n \neq n') \,,  \\
\int\limits_{S^2} h_{nn'}(\Omega,\Omega') 
\,\rho^{(1)}(\Omega') \, d\Omega' & = 0 \quad (n \neq n') \,.
\end{align}
\end{subequations}
Last, in the asymptotic limit of large particle separations,
$g_{nn'}(\Omega,\Omega')$ and $h_{nn'}(\Omega,\Omega')$ behave like
\begin{align}
\label{eq16}
\lim_{|{\bf x}_{nn'}| \rightarrow \infty} g_{nn'}(\Omega,\Omega') &=
1 \,, \\
\label{eq17}
\lim_{|{\bf x}_{nn'}| \rightarrow \infty} h_{nn'}(\Omega,\Omega') &=
0 \,,
\end{align}
independent of the direction of ${\bf x}_{nn'}$. This follows due to
$\lim_{|{\bf x}_{nn'}| \rightarrow \infty} 
\rho^{(2)}_{nn'}(\Omega,\Omega') \rightarrow \rho^{(1)}(\Omega)\,
\rho^{(1)}(\Omega')$, in full agreement with the behaviour of a liquid
system. 

Making use of Eqs.~(\ref{eq10}),~(\ref{eq12}),~(\ref{eq13}) and
introducing 
\begin{equation}
\label{eq18}
D(\Omega,\Omega') = 4\pi \left(
 \rho^{(1)}(\Omega)\,\delta(\Omega|\Omega')-
 \rho^{(1)}(\Omega) \, \rho^{(1)}(\Omega') \right) 
\end{equation}
we can rewrite $G^{(\alpha)}_{nn'}(\Omega,\Omega')$, $\alpha = s,d$, as follows:
\begin{subequations}
\label{eq19}
\begin{align}
\label{eq19a}
G^{(s)}_{nn'}(\Omega,\Omega') & = \frac{1}{4\pi} \, \delta_{nn'} 
D(\Omega,\Omega') \,, \\
\label{eq19b}
G^{(d)}_{nn'}(\Omega,\Omega') &= \left( 1 - \delta_{nn'} \right)
\rho^{(1)}(\Omega) \, h_{nn'}(\Omega,\Omega')\, \rho^{(1)}(\Omega') \,.
\end{align}
\end{subequations}

As for molecular liquids~\cite{GG1984} (see also,
e.g.,~\cite{RSS1994,LL1999,SS1997,FFG1997}) 
it is useful to expand all orientation-dependent functions
with respect to a complete set of functions,
determined by the rotational symmetry. If $\mathcal{P}$ is the point
symmetry group of the lattice and $\mathcal{P}_{M}$ the symmetry group
of the molecules, one can use basis functions for irreducible
representations of the symmetry group of $\rho^{(1)}(\Omega)$, which
is a subgroup of $\mathcal{P} \otimes
\mathcal{P}_{M}$~\cite{LM1994,YP1980,BP1988}. For axially symmetric
particles, these are linear combinations of the spherical harmonics 
$Y_{\lambda}(\Omega)$, $\lambda = (lm)$~\cite{LM1994,YP1980}. On the other
hand, the spherical harmonics itself can be taken. We have chosen the 
latter possibility in order to keep similarity to molecular liquids. 
Consequently, we have for any functions $f(\Omega)$ and
$F_{nn'}(\Omega,\Omega')$ their $\lambda$-transforms 
and the corresponding inverse transformations:
\begin{subequations}
\label{eq20}
\begin{eqnarray}
\label{eq20a}
f_{\lambda} & = & i^{-l}\int\limits_{S^2} f(\Omega) \,
Y_{\lambda}^{*}(\Omega) \,d \Omega, \\
\label{eq20b}
f(\Omega) & = & \sum_{\lambda} \, (-i)^{-l} f_{\lambda}
\,Y_{\lambda}(\Omega), 
\end{eqnarray}
\end{subequations}
\begin{subequations}
\label{eq21}
\begin{align}
\label{eq21a}
F_{nn',\lambda\lambda'} = i^{l'-l}&\iint\limits_{S^{2}S^{2}}
F_{nn'}(\Omega,\Omega') \, Y_{\lambda}^{*}(\Omega)
\, Y_{\lambda'} (\Omega') \, d \Omega \, d \Omega', \quad\quad\\
\label{eq21b}
F_{nn'}(\Omega,\Omega') & = \sum_{\lambda\lambda'} \,
(-i)^{l'-l} F_{nn',\lambda\lambda'} \, Y_{\lambda}(\Omega) 
\,Y^{*}_{\lambda'}(\Omega') .
\end{align}
\end{subequations}
The purely imaginary prefactors in~(\ref{eq20a}) and~(\ref{eq21a}) are
taken for technical convenience. 

Finally, we can use the lattice Fourier transform due to the lattice
translational invariance. It is restricted
to the first Brillouin zone of volume $V_{\text{BZ}}$. For example, the
transform of the site-site matrix elements~(\ref{eq21a}) and its 
inverse are given
\begin{subequations}
\label{eq22}
\begin{eqnarray}
\label{eq22a}
F_{\lambda\lambda'}({\bf q}) & = & \sum\limits_{{\bf x}_{nn'}}
\text{e}^{i{\bf q}\cdot {\bf x}_{nn'}} F_{nn',\lambda\lambda'}, \\
\label{eq22b}
F_{nn',\lambda\lambda'} & = & \frac{1}{V_{\text{BZ}}}
\int\limits_{\text{1.BZ}} F_{\lambda\lambda'}({\bf q})\,
\text{e}^{-i {\bf q}\cdot {\bf x}_{nn'}} d^{3} q.
\end{eqnarray}
\end{subequations}
\subsection{\label{secIIb}Symmetry relations for the one-particle distribution and
the correlation functions}
Let us first discuss the one-particle orientational distribution
function $\rho^{(1)}(\Omega)$. $\rho^{(1)}(\Omega)$ must carry the
full point symmetry $\mathcal{P}$ of the underlying periodic lattice, 
and also the symmetry $\mathcal{P}_{M}$ of the particles~\cite{LM1994,YP1980}. 
Neglecting the latter for a moment, $\rho^{(1)}(\Omega)$ for axially
symmetric particles can be expanded into a series of all 
$\mathcal{P}$-invariant combinations $\hat{Y}_{ln_{l}}(\Omega)$
of spherical harmonics. This expansion reads (cf.~(\ref{eq20b}))
\begin{equation}
\label{eq23}
\rho^{(1)}(\Omega) = \sum\limits_{ln_{l}}  \, (-i)^{-l}
\,\rho^{(1)}_{ln_{l}} \,\hat{Y}_{ln_{l}}(\Omega)\,.
\end{equation}
Here, the numbers 
$n_{l}$ are the multiplicities of the unity irreducible representation
contained in the point group representation of $\mathcal{P}$
established by all spherical harmonics of order $l$. Then, by the 
invariance requirement under symmetry operations of the particles,~(\ref{eq23}) can 
eventually be further simplified~\cite{LM1994,YP1980}.

If the lattice is cubic, $\mathcal{P} = O_{h}$, and for any kind of 
axially symmetric particles we have
\begin{eqnarray}
\label{eq24}
\rho^{(1)}(\Omega) =  &\frac{1}{\sqrt{4\pi}} \hat{Y}_{01}(\Omega) + 
	\rho^{(1)}_{41} \,\hat{Y}_{41}(\Omega) -
	\rho^{(1)}_{61} \,\hat{Y}_{61}(\Omega) \nonumber \\
	& + \rho^{(1)}_{81} \,\hat{Y}_{81}(\Omega) + O(l = 10).
\end{eqnarray}
The $O_{h}$ cubic invariants $\hat{Y}_{01}(\Omega) =
Y_{00}(\Omega) = (4\pi)^{-1/2}$, $\hat{Y}_{41}(\Omega)$,
$\hat{Y}_{61}(\Omega)$ and $\hat{Y}_{81}(\Omega)$ are 
real functions and given in \cite{LB1947},
\footnote{To our opinion, the factor $\frac{1}{6}$ in
the expression for $\alpha_{8}$ in \cite{LB1947} is a misprint and 
should read $\frac{1}{3}$ instead.} up to factors $(4\pi)^{-1/2}
\rho^{-l}$. 

In Sec.~\ref{secIV}, the canonical averages 
$\langle Y_{\lambda}\rangle = \int_{S^2} \rho^{(1)}(\Omega) \,Y_{\lambda} (\Omega)
\,d\Omega$ are needed. The values $\langle \hat{Y}_{41}\rangle$, 
$\langle \hat{Y}_{61}\rangle$ and $\langle \hat{Y}_{81}\rangle$ 
have been calculated by MC simulations (see Sec.~\ref{secIV}) for several
values of $a$ and $b$, and the nonvanishing $\langle Y_\lambda\rangle$ 
up to $l = 9$ are
\begin{align}
\label{eq25}
\langle Y_{00}\rangle &=\frac{1}{\sqrt{4\pi}}, &
\langle Y_{40}\rangle &= \frac{\sqrt{21}}{6} \langle \hat{Y}_{41}\rangle, \nonumber\\
\langle Y_{4\pm4}\rangle &= \frac{\sqrt{30}}{12} \langle \hat{Y}_{41}\rangle, &
\langle Y_{60}\rangle &= \frac{\sqrt{2}}{4} \langle \hat{Y}_{61}\rangle, \nonumber\\
\langle Y_{6\pm4}\rangle &= -\frac{\sqrt{7}}{4} \langle \hat{Y}_{61}\rangle, &
\langle Y_{80}\rangle &= \frac{\sqrt{33}}{8} \langle \hat{Y}_{81}\rangle,\nonumber\\
\langle Y_{8\pm4}\rangle &= \frac{\sqrt{42}}{24} \langle \hat{Y}_{81}\rangle, &
\langle Y_{8\pm8}\rangle &= \frac{\sqrt{390}}{48} \langle \hat{Y}_{81}\rangle. 
\end{align}

Next, we investigate the symmetries of $G_{nn'}(\Omega,\Omega')$. From
the definition~(\ref{eq8}) it follows immediately 
\begin{subequations}
\label{eq26}
\begin{equation}
\label{eq26a}
G_{nn'}(\Omega,\Omega') = G_{n'n}(\Omega',\Omega).
\end{equation}
If the inversion $I$ belongs to $\mathcal{P}$, the inverted 
$G_{nn'}(\Omega,\Omega')$ must match the old one by use of 
Eq.~(\ref{eq8}), since $H(\{\Omega_{n} \}, \{ {\bf l}_{n} \})$ 
remains unchanged under inversion then:
\begin{equation}
\label{eq26b}
G_{n'n}(-\Omega,-\Omega') = G_{nn'}(\Omega,\Omega'),
\end{equation}
where $\Omega = (\theta,\phi) \longleftrightarrow -\Omega \equiv I \Omega = 
(\pi-\theta,\phi+\pi)$  has been used. If the symmetry group 
$\mathcal{P}$ contains rotations $R$, the rotated
$G_{nn'}(\Omega,\Omega')$ must be the original one:
\begin{equation}
\label{eq26c}
G_{R{\bf x}_{nn'}}(R\Omega,R\Omega') = 
G_{{\bf x}_{nn'}}(\Omega,\Omega').
\end{equation}
\end{subequations}
The knowlegde of the transforms on the lhs of~(\ref{eq26b}) 
and~(\ref{eq26c}) is sufficient to determine the transform 
$G_{g{\bf x}_{nn'}}(g\Omega,g\Omega')$ for an arbitrary point symmetry 
operation $g$, which must be $G_{{\bf x}_{nn'}}(\Omega,\Omega')$ if
$g \in \mathcal{P}$. 

The properties~(\ref{eq26}) affect the matrix elements
$G_{nn',\lambda\lambda'}$ the following way:
\begin{subequations}
\label{eq27}
\begin{align}
\label{eq27a}
G_{nn',\lambda\lambda'} & =
G^{*}_{n'n,\lambda'\lambda}\,, \\
\label{eq27b}
G_{nn',\lambda\lambda'} & = (-1)^{l+l'}
G_{n'n,\lambda\lambda'}\,, \\
\label{eq27c}
G_{R{\bf x}_{nn'},lm,l'm'} & = \sum\limits_{m''m'''} 
D^{l}_{mm''}(R) \, D^{l'^{*}}_{m'm'''}(R)\nonumber\\
&\quad\quad\quad\quad\quad \times G_{{\bf x}_{nn'},lm'',l'm'''}\,.
\end{align}
In~(\ref{eq27c}), Wigner's generalized spherical harmonics
(rotation matrices) are used~\cite{GG1984}. To calculate the correct 
rotation matrix elements for~(\ref{eq27c}), one uses the three Euler angles
carrying some coordinate frame, in which ${\bf x}_{nn'}$ is assumed
to be fixed, into a new, symmetry-equivalent one such that the 
rotated vector ${\bf x}_{nn'}$ coincides with $R{\bf x}_{nn'}$.
The behaviour of $\{Y_{lm}(\Omega)\}$ under complex conjugation yields a fourth
property:
\begin{equation}
\label{eq27d}
G_{nn',lm,l'm'} = (-1)^{l+l'+m+m'} G^{*}_{nn',l-m.l'-m'}
\end{equation}
\end{subequations}
Eqs.~(\ref{eq27}) are translated to the Fourier transformed matrices 
${\bf S}({\bf q}) = 4\pi \,{\bf G}({\bf q}) = 4 \pi \left(
G_{\lambda\lambda'} ({\bf q}) \right)$ (see Eqs.~(\ref{eq40}),~(\ref{eq41}))
due to
\begin{subequations}
\label{eq28}
\begin{align}
\label{eq28a}
S_{\lambda\lambda'}({\bf q}) & =
S^{*}_{\lambda'\lambda}({\bf q}) \,, \\
\label{eq28b}
S_{\lambda\lambda'}({\bf q}) & = (-1)^{l+l'}
S_{\lambda\lambda'}(-{\bf q})\,, \\
\label{eq28c}
S_{lm,l'm'}(R{\bf q}) & = \sum\limits_{m''m'''} 
D^{l}_{mm''}(R) \, D^{l'^{*}}_{m'm'''}(R)\nonumber\\
&\quad\quad\quad\quad\quad \times S_{lm'',l'm'''}({\bf q})\,,\\
\label{eq28d}
S_{lm,l'm'}({\bf q}) & = (-1)^{l+l'+m+m'} 
S^{*}_{l-m.l'-m'}(-{\bf q})
\end{align}
\end{subequations}
(\ref{eq28a}) demonstrates that ${\bf S}({\bf q})$ is
Hermitian, but ${\bf G}_{nn'}$ in general is not.

The symmetry of the particles can bring about extra
characteristics of the matrix elements~\cite{LM1994,YP1980}. 
For axially symmetric particles with inversion symmetry,
$G^{(d)}_{nn'}(\Omega,\Omega') = G^{(d)}_{nn'}(-\Omega,\Omega') =
G^{(d)}_{nn'}(\Omega,-\Omega') = G^{(d)}_{nn'}(-\Omega,-\Omega')$
is valid, and the self part fulfills
$G^{(s)}_{nn}(\Omega,\Omega') = G^{(s)}_{nn}(-\Omega,-\Omega')$.
Consequently, ${\bf G}^{(s)}_{nn}$ can have nonzero elements for $l$
and $l'$ even or $l$ and $l'$ odd, respectively, while ${\bf G}^{(d)}_{nn'}$ has 
nontrivial elements for $l$ and $l'$ even, only.
  
We want to conclude this section with the remark for cubic lattices, 
that by symmetry the knowledge of the correlations for only $\frac{1}{48}$ of 
all lattice vectors or $\frac{1}{48}$ of the volume of the first
Brillouin zone is necessary to calculate the correlations for the
complete lattice or Brillouin zone.
\section{\label{secIII}Ornstein-Zernike equation and Percus-Yevick
approximation}
Similarly to simple and molecular liquids, we will introduce the direct
correlation function $c_{nn'}(\Omega,\Omega')$, which is related to 
$h_{nn'}(\Omega,\Omega')$ by the OZ
equation~\cite{HMD1990,GG1984}. Since $c_{nn'}(\Omega,\Omega')$ is
determined by the inverse functions of $G^{(s)}_{nn'}(\Omega,\Omega')$ and 
$G_{nn'}(\Omega,\Omega')$, one has to be careful because of
relations~(\ref{eq11}), which imply that a constant function
$f(\Omega) = const$ is an eigenfunction of
$G^{(s)}_{nn'}(\Omega,\Omega')$ and 
$G_{nn'}(\Omega,\Omega')$ with eigenvalue zero. Therefore, 
these inverses do not exist on the
one-dimensional subspace of constant functions. This is a new feature
occuring for rigid molecular crystals. The solution of this problem is
easy using the $\lambda$-transforms $\frac{1}{4\pi} \delta_{nn'}
D_{\lambda\lambda'}$ and $G_{nn',\lambda\lambda'}$ of 
$G^{(s)}_{nn'}(\Omega,\Omega')$ and $G_{nn'}(\Omega,\Omega')$,
respectively. Because of Eqs.~(\ref{eq11}), it follows that the first
rows and columns of the matrices ${\bf D}$ and ${\bf G}_{nn'}$ are zero,
i.e.
\begin{equation}
\label{eq29}
D_{\lambda\lambda'} = 0\,, \quad G_{nn',\lambda\lambda'} = 0\,,\quad
l = 0 \text{ and/or } l' =0\,.
\end{equation}
This is reasonable since quantities with $l=0$ and/or $l'=0$ do not
describe orientational degrees of freedom and therefore are
unphysical. Obviously inversion must be restricted to the
$\lambda$-transform with $ll' > 0$, which is equivalent to restrict
inversion of $G^{(s)}_{nn'}(\Omega,\Omega')$ and
$G_{nn'}(\Omega,\Omega')$ to the space of functions having no constant
part. Since an arbitrary function $F_{nn'}(\Omega,\Omega')$, and maybe 
$c_{nn'}(\Omega,\Omega')$ as well, need not fulfill
relations~(\ref{eq11}), we associate with any function 
$F_{nn'}(\Omega,\Omega')$ on $S^{2} \times S^{2}$ the function
\begin{widetext}
\begin{subequations}
\label{eq30}
\begin{equation}
\label{eq30a}
F_{nn'}^{\circ}(\Omega,\Omega') = F_{nn'} (\Omega,\Omega')
-\frac{1}{4\pi}\int\limits_{S^{2}} F_{nn'}(\Omega,\Omega')\,d \Omega 
-\frac{1}{4\pi}\int\limits_{S^{2}} F_{nn'}(\Omega,\Omega')\,d \Omega'
+\frac{1}{(4\pi)^{2}}\iint\limits_{S^{2}S^{2}} F_{nn'}
(\Omega,\Omega')\,d \Omega \,d \Omega',
\end{equation}
which can be rewritten as
\begin{equation}
\label{eq30b}
F_{nn'}^{\circ}(\Omega,\Omega') = 
\iint\limits_{S^{2}S^{2}} \left( \delta
(\Omega|\Omega'') - \frac{1}{4\pi} \right) 
F_{nn'}(\Omega'',\Omega''')
\left( \delta (\Omega'''|\Omega') - \frac{1}{4\pi}\right)
d\Omega''d\Omega'''.
\end{equation}
\end{subequations}
\end{widetext}
It is important to note that the two functions $F_{nn'}(\Omega,\Omega')$
and $F_{nn'}^{\circ}(\Omega,\Omega')$ differ at most by the sum of a
function of $\Omega$ only, a function of $\Omega'$ only and a constant 
function, and that $F_{nn'}^{\circ}(\Omega,\Omega')$ fulfills
relations~(\ref{eq11}). In general, $F_{nn'}(\Omega,\Omega')$ cannot be
recalculated from $F_{nn'}^{\circ}(\Omega,\Omega')$,
since~(\ref{eq30}) describes a projection. The associated
projection operator $R = R^{2} = R^{\dagger}$ is
\begin{subequations}
\label{eq31}
\begin{equation}
R(\Omega,\Omega')  =
\delta (\Omega|\Omega') - \frac{1}{4\pi} \equiv \delta^{\circ}
(\Omega|\Omega') \,.
\end{equation}
\end{subequations}
By Eq.~(\ref{eq30b}), we can present $F_{nn'}^{\circ}(\Omega,\Omega')$
in the shorthand notation 
\begin{subequations}
\label{eq32}
\begin{equation}
\label{eq32a}
F^{\circ}_{nn'}(\Omega,\Omega')  =
(RF_{nn'}R)(\Omega,\Omega')\,.
\end{equation}
\end{subequations}
The elimination of the first column and row (cf. Eq.~(\ref{eq29})) of
the $\lambda$-transform $F_{nn',\lambda\lambda'}$ can be done
similarly by use of the $\lambda$-transform
\begin{equation}
\tag{\ref{eq31}b}
\label{eq31b}
R_{\lambda\lambda'} =  i^{l'-l} \left( \delta_{\lambda\lambda'} -
 \delta_{\lambda,00} \delta_{00,\lambda'} \right)
\end{equation}
of $R(\Omega,\Omega')$. This leads to 
\begin{equation}
\tag{\ref{eq32}b}
\label{eq32b}
F^{\circ}_{nn',\lambda\lambda'} = \sum\limits_{\lambda''\lambda'''}
R_{\lambda\lambda''}  F_{nn',\lambda''\lambda'''}  R_{\lambda'''\lambda'}
\end{equation}
Relations~(\ref{eq11}) imply that 
\begin{equation}
\label{eq33}
G_{nn'}(\Omega,\Omega') \equiv G^{\circ}_{nn'}(\Omega,\Omega') \,,\;
D(\Omega,\Omega') \equiv D^{\circ}(\Omega,\Omega')
\end{equation}
for all $nn'$, but, in general,
\begin{equation}
\label{eq34}
h^{\circ}_{nn'}(\Omega,\Omega') \neq h_{nn'}(\Omega,\Omega')
\quad (n \neq n')\,.
\end{equation}

For later purposes it will be helpful to rewrite Eq.~(\ref{eq19b})
replacing $h_{nn'}(\Omega,\Omega')$ by $h^{\circ}_{nn'}(\Omega,\Omega')$.
Eqs.~(\ref{eq15}) can be used to show that the rhs of 
Eq.~(\ref{eq19b}) is indeed of the form $\circ$ and that the function
$D(\Omega,\Omega')$ can be introduced as follows:
\begin{align}
\label{eq38}
\rho^{(1)}(\Omega)\,h_{nn'}(\Omega,\Omega') \,\rho^{(1)}&(\Omega') 
\nonumber \\
 = \frac{1}{(4\pi)^2}\iint\limits_{S^{2}S^{2}} D(\Omega,\Omega'')\,
&h_{nn'}(\Omega'',\Omega''')\,
D(\Omega''',\Omega')\,d\Omega''\,d\Omega''' \nonumber \\
 = \frac{1}{(4\pi)^2}\iint\limits_{S^{2}S^{2}} D(\Omega,\Omega'')\,
&h^{\circ}_{nn'}(\Omega'',\Omega''')\, 
D(\Omega''',\Omega')\,d\Omega''\,d\Omega''' \nonumber \\
& (n \neq n') \,,
\end{align}
where the last line involves $h^{\circ}_{nn'}(\Omega,\Omega')$ instead of
$h_{nn'}(\Omega,\Omega')$ since $D(\Omega,\Omega') = (RD)(\Omega,\Omega') =
(DR)(\Omega,\Omega') = (RDR)(\Omega,\Omega')$ projects out
the constant parts of $h_{nn'}(\Omega,\Omega')$. Contrary to the general case, it
will be shown in App.~\ref{appI} that 
$h^{\circ}_{nn'}(\Omega,\Omega')$ already determines
$h_{nn'}(\Omega,\Omega')$ uniquely (cf. Eq.~(\ref{eqA4})). 
The last step is just to write down
$G^{\circ}_{nn'}(\Omega,\Omega')$ in the new form
\begin{align}
\label{eq39}
& G^{\circ}_{nn'}(\Omega,\Omega') = \frac{1}{4\pi} \delta_{nn'}
D(\Omega,\Omega') + \frac{1}{(4\pi)^2}\left(1
-\delta_{nn'} \right) \nonumber \\
\times & \iint\limits_{S^{2}S^{2}} D(\Omega,\Omega'')\,
h^{\circ}_{nn'}(\Omega'',\Omega''')\, D(\Omega''',\Omega')
\,d\Omega''\,d\Omega'''
\end{align}

The tensorial static structure factors
$S_{\lambda\lambda'}({\bf q}) \equiv S^{\circ}_{\lambda\lambda'}({\bf
q})$ also have their first columns ($l'= 0$) and rows ($l =0$) vanishing and 
can be defined by the two equivalent expressions
\begin{align}
\label{eq40}
S_{\lambda\lambda'}({\bf q}) & = \frac{4\pi}{N} \,\langle 
\delta\rho^{*}_{\lambda}({\bf q})\,
\delta\rho_{\lambda'}({\bf q}) \rangle
 = 4\pi \, G_{\lambda\lambda'}({\bf q}) \,,
\end{align}
where the factor $4 \pi$ is introduced as 
the inverse of the direction average of the one-particle density 
$\rho^{(1)}(\Omega)$. The transform of
Eq.~(\ref{eq39}) leads directly to the concise expression 
for the structure factors $S_{\lambda\lambda'}({\bf q})$ in matrix form,
\begin{equation}
\label{eq41}
{\bf S} ({\bf q})  = {\bf D} + \frac{1}{4\pi} 
{\bf D} \,{\bf h}^{\circ} ({\bf q})\,{\bf D}\,, 
\end{equation}
where for the Fourier transform ${\bf h}^{\circ} ({\bf q})$ 
it is assumed that $h_{nn}(\Omega,\Omega') \equiv 0$.
Eq.~(\ref{eq41}) closely resembles the corresponding relations for 
simple~\cite{HMD1990} and molecular liquids~\cite{GG1984}.

There is a second problem occuring if hard core interactions lead to $K
\subset S^{2}$. As already discussed in Sec.~\ref{secIIa},
$\rho^{(1)}(\Omega)$ and therefore $G_{nn'}(\Omega,\Omega')$ vanish
in that case on the subset $\bar{K} \subset S^{2}$ for all $nn'$. Therefore, any
function $g(\Omega)$ which vanishes on $K$ but is nonzero on the
complement $\bar{K}$ is an eigenfunction of $G_{nn'}(\Omega,\Omega')$
with eigenvalue zero, too. This problem can be properly solved as
well. It is reasonable to discuss hard core interactions leading to
$K \subset S^{2}$ separately. Therefore, the next
subsection~\ref{secIIIa} will present the OZ equation and the PY
approximation for $K = S^{2}$, and in subsection~\ref{secIIIb} systems 
with $K \subset S^{2}$ will be treated.
\subsection{\label{secIIIa}Molecular crystals characterized by $K = S^{2}$}
Of course, $K = S^{2}$ for soft potentials. But also for hard
core interactions it can be $K = S^{2}$, provided size and
shape of the particles are properly chosen. This can be the case, e.g., for
slightly aspherical hard particles.

Above, everything has been reduced to functions of the form $\circ$, and
also the direct correlation function can only be defined as 
$c^{\circ}_{nn'}(\Omega,\Omega')$, at least at this first stage of definition:
\begin{align}
\label{eq42}
&c^{\circ}_{nn'}(\Omega,\Omega') = \nonumber \\
&\left(\left(G^{(s)^{\circ}}\right)^{-1}\right)^
{\circ}_{nn'}(\Omega,\Omega') - \left(\left(G^{\circ}\right)^{-1}\right)^
{\circ}_{nn'}(\Omega,\Omega')
\end{align}
The reader should note that, in contrast to $\rho^{(2)}_{nn'}(\Omega,\Omega')$ and 
$h^{\circ}_{nn'}(\Omega,\Omega')$, the self part of the direct correlation function
exists and is well defined by Eq.~(\ref{eq42}).

The function $G^{\circ}_{nn'}(\Omega,\Omega')$ has an inverse on
the subspace of functions $g^{\circ}(\Omega)$ only, since
$G_{nn'}(\Omega,\Omega') \equiv G^{\circ}_{nn'}(\Omega,\Omega') = (R
G_{nn'} R) (\Omega,\Omega')$ maps constant functions to zero on $S^{2}$. 
This inverse is determined itself only in the form $\circ$ and fulfills
\begin{align}
\label{eq43}
\sum\limits_{n''} \int\limits_{S^{2}} &
\left(\left(G^{\circ}\right)^{-1}\right)^
{\circ}_{nn''}(\Omega,\Omega'') \,G^{\circ}_{n''n'}
(\Omega'',\Omega') \,d\Omega'' \nonumber \\
& =
\delta_{nn'} R (\Omega,\Omega')\, , 
\quad \Omega,\Omega' \in S^{2}
\end{align}
Note that the rhs of Eq.~(\ref{eq43}) does not involve
$\delta (\Omega|\Omega')$ but $R(\Omega|\Omega') \equiv 
\delta^{\circ} (\Omega|\Omega')$. 
\begin{widetext}
An analogous equation defines the inverse of the
self part Eq.~(\ref{eq10a}), which is, contrary to the full inverse
above, known explicitely:
\begin{equation}
\label{eq44}
\left(\left(G^{(s)^{\circ}}\right)^{-1}\right)^
{\circ}_{nn'}(\Omega,\Omega') = \delta_{nn'}\Bigg( 
\frac{\delta(\Omega|\Omega')}{\rho^{(1)}(\Omega)} 
- \frac{1}{4\pi} \frac{1}{\rho^{(1)}(\Omega)} -
\frac{1}{4\pi} \frac{1}{\rho^{(1)}(\Omega')}
+\frac{1}{(4\pi)^{2}} \int\limits_{S^{2}} \frac{1}{\rho^{(1)}(\Omega)} \, d\Omega
\Bigg) \, .
\end{equation}

The derivation of the lattice OZ equation causes no problems having
the tools presented in Secs.~\ref{secII} and~\ref{secIII} above. 
The procedure is to substitute
$((G^{\circ})^{-1})^{\circ}_{nn'}(\Omega,\Omega')$ 
from  Eq.~(\ref{eq42}) and
$G^{\circ}_{nn'}(\Omega,\Omega')$ from
Eq.~(\ref{eq39}) into Eq.~(\ref{eq43}), at the same time
using that $G^{(s)}_{nn'}(\Omega,\Omega') = \frac{1}{4\pi}\,\delta_{nn'}D(\Omega,\Omega')$ and
$\int_{S^{2}} ((D^{\circ})^{-1})^{\circ}(\Omega,\Omega'') \,D(\Omega'',\Omega')
\, d\Omega'' = R(\Omega,\Omega')$. Then, the lhs of Eq.~(\ref{eq43}) becomes
\begin{align}
\label{eq45}
&\sum\limits_{n''} \int\limits_{S^{2}}
\left( 4\pi \, \delta_{nn''}\left(\left(D^{\circ}\right)^{-1}\right)^{\circ}
(\Omega,\Omega'')-c^{\circ}_{nn''}(\Omega,\Omega'')\right)
\nonumber \\
&\left( \frac{1}{4\pi} \delta_{n''n'}
D(\Omega'',\Omega') + \frac{1}{(4\pi)^2}\left(1
-\delta_{n''n'} \right) \iint\limits_{S^{2}S^{2}} D(\Omega'',\Omega''')\,
h^{\circ}_{n''n'}(\Omega''',\Omega'''')\, D(\Omega'''',\Omega')
\,d\Omega'''\,d\Omega'''' \right) d\Omega''
\nonumber \\
& = \delta_{nn'} R(\Omega,\Omega') 
- \frac{1}{4\pi} \int\limits_{S^{2}} c^{\circ}_{nn'}(\Omega,\Omega'')
\, D(\Omega'',\Omega') \, d \Omega''
+  \frac{1}{4\pi} \left(1-\delta_{nn'}\right) \int\limits_{S^{2}} h^{\circ}_{nn'}
(\Omega,\Omega'')\, D(\Omega'',\Omega') \, d \Omega'' 
\nonumber\\
& - \frac{1}{(4\pi)^2} \sum\limits_{n''}\left(1-\delta_{n''n'}\right)
\iiint\limits_{S^{2}S^{2}S^{2}} c^{\circ}_{nn''}(\Omega,\Omega'')\,
D(\Omega'',\Omega''') \, h^{\circ}_{n''n'}(\Omega''',\Omega'''') 
\,D(\Omega'''',\Omega')\,d \Omega''\,d \Omega'''\,d \Omega''''
\end{align}
for $\Omega,\Omega' \in S^{2}$, which must match the rhs of
Eq.~(\ref{eq43}), i.e. $\delta_{nn'} R(\Omega,\Omega')$. 
$\delta_{nn'} R(\Omega,\Omega')$ cancels. Operating with
$\int_{S^{2}} \ldots\, ((D^{\circ})^{-1})^{\circ} 
(\Omega',\Omega''''') \, d\Omega'$ on both sides gives the OZ equation 
for the lattice in the site-site angular representation
for $\Omega,\Omega' \in S^{2}$:
\begin{subequations}
\label{eq46}
\begin{align}
\label{eq46a}
\left(1-\delta_{nn'}\right) h^{\circ}_{nn'}
(\Omega,\Omega') =  c^{\circ}_{nn'}(\Omega,\Omega') + 
\frac{1}{4\pi} \sum\limits_{n'' \neq n'}
\iint\limits_{S^{2}S^{2}} c^{\circ}_{nn''}(\Omega,\Omega'')\,
D(\Omega'',\Omega''') \, h^{\circ}_{n''n'}(\Omega''',\Omega') 
\,d \Omega''\,d \Omega'''\,.
\end{align}
\end{subequations}
\end{widetext}
The associated $\lambda$- and Fourier transformed matrix equation, which has the most
significant form and which is used for the numerical work, reads
\begin{equation}
\label{eq46b}
\tag{\ref{eq46}b}
{\bf h}^{\circ}({\bf q}) = {\bf c}^{\circ}({\bf q}) + 
\frac{1}{4\pi} {\bf c}^{\circ}({\bf q})\,{\bf D}\,{\bf
h}^{\circ}({\bf q}) \,, 
\end{equation}
where again it is assumed that $h_{nn}(\Omega,\Omega') \equiv 0$.

This result is almost the same as for molecular
liquids~\cite{GG1984}. There are two main differences. First, we have
to use all matrices in their form $\circ$, i.e. the first row and
column of the original matrices are skipped, since they are
zero. Second, there appears the matrix ${\bf D}$, which is 
identical to ${\bf G}^{(s)}$, apart from a prefactor, and third the
Fourier backtransform of ${\bf h}^{\circ}({\bf q})$ has to fulfill 
$h_{nn,\lambda\lambda'} \equiv 0$ for all $\lambda\lambda'$. 

Up to now, the equations are not closed. Since neither the total 
correlation nor the direct correlation function is given, the
previous concepts are almost useless if one is interested in an
analytical approach to determine the structure factors~(\ref{eq40}),~(\ref{eq41}). 
An additional equation, called the closure relation, 
must be used to find a self-consistent solution for $h_{nn'}(\Omega,\Omega')$
and $c_{nn'}(\Omega,\Omega')$, as for simple and molecular
liquids. It has been our intention to follow as close as possible 
the established lines of liquid theory, and so we chose the most
straightforward analogon of the PY approximation for the lattice,
which is for $n \neq n'$
\begin{align}
c_{nn'}(\Omega,\Omega') &=
f_{nn'}(\Omega,\Omega') \left(g_{nn'}(\Omega,\Omega') - 
c_{nn'}(\Omega,\Omega') \right) \nonumber \\
\label{eq47}
&= f_{nn'}(\Omega,\Omega') \left(1+h_{nn'}(\Omega,\Omega') - 
c_{nn'}(\Omega,\Omega') \right).
\end{align}
Note that Eq.~(\ref{eq42}) involves $c^{\circ}_{nn'}(\Omega,\Omega')$, only,
but in~(\ref{eq47}) the full function 
$c_{nn'}(\Omega,\Omega')$ appears. In App.~\ref{appIV}, it will be
shown for hard particles and $n \neq n'$ that $c^{\circ}_{nn'}(\Omega,\Omega')$
determines $c_{nn'}(\Omega,\Omega')$ uniquely.
$f_{nn'}(\Omega,\Omega')$ is the 
Mayer $f$-function, which is for $n \neq n'$
\begin{equation}
\label{eq48}
f_{nn'}(\Omega,\Omega') = \exp\{-\beta
V_{nn'}(\Omega_{n},\Omega_{n'})\} -1\,.
\end{equation}

For hard particles, the pair potential is $V_{nn'}(\Omega,\Omega') =
0$ (if the pair $(n\Omega,n'\Omega')$ has no overlap) or 
$V_{nn'}(\Omega,\Omega') = \infty$ (if the pair has overlap).
This implies that all static properties are athermal and that it is
\begin{align}
\label{eq49}
f_{nn'}(\Omega,\Omega') &= 
\left\{
\begin{array}{rc}
0\,, & \text{no overlap}\,, \\
-1 \,, & \text{overlap}\,.
\end{array}
\right .
\end{align}
The range of the $f$-function for hard ellipsoids is
$\text{max} (a,b)$, if the ellipsoids are fixed with 
their centers of mass on the lattice. 
In accordance with the theory of liquids of hard
particles,~(\ref{eq47}) yields $g_{nn'}(\Omega,\Omega')
= 0$ $\longleftrightarrow$ $\rho^{(2)}_{nn'}(\Omega,\Omega') = 0$
while $c_{nn'}(\Omega,\Omega')$ remains undetermined, if 
$(n\Omega,n'\Omega')$ has overlap and $c_{nn'}(\Omega,\Omega') = 0$ 
while $g_{nn'}(\Omega,\Omega')$ remains undetermined, if 
$(n\Omega,n'\Omega')$ has no overlap~\cite{HMD1990,GG1984}.

\begin{widetext}
Eq.~(\ref{eq47}) should be available in its $\lambda$-transform, since the
numerical solution of the OZ equation using the PY approximation is
most conveniently done in terms of matrix elements. From
Eq.~(\ref{eq47}), it is straightforward to deduce that
\begin{align}
\label{eq50}
&c_{nn',\lambda\lambda'} = \sum\limits_{\lambda''}
\sum\limits_{\lambda'''} \sum\limits_{\lambda''''}
\sum\limits_{\lambda'''''} \,i^{l'-l+l''-l'''+l''''-l'''''} \,
\left[ \frac{(2l''+1)(2l''''+1)}{4\pi \,(2l+1)}
\right]^{\frac{1}{2}}
\left[ \frac{(2l'''+1)(2l'''''+1)}{4\pi \,(2l'+1)}
\right]^{\frac{1}{2}} \nonumber\\
&C(l''l''''l,000) \, C(l'''l'''''l',000) \,
C(l''l''''l,m''m''''m) \, C(l'l'''l''''',m'''m'''''m')\,
f_{nn',\lambda''\lambda'''}
\left(g_{nn',\lambda''''\lambda'''''}
-c_{nn',\lambda''''\lambda'''''}\right),
\end{align}
where the Clebsch Gordan coefficients
$C(l_{1}l_{2}l_{3},m_{1}m_{2}m_{3})$ are given in~\cite{GG1984}. 
\end{widetext}

The numerical use of Eq.~(\ref{eq50}) requires the knowledge of the
matrix elements $f_{nn',\lambda\lambda'}$ and also 
$g_{nn',\lambda\lambda'}$ or $h_{nn',\lambda\lambda'}$, $n \neq
n'$. These are derived in Apps.~\ref{appII} and~\ref{appIII}. Having
calculated the matrix ${\bf c}_{nn'}$ then, the Fourier transform of 
${\bf c}^{\circ}_{nn'} = {\bf R} \,{\bf c}_{nn'} \, 
{\bf R}$ serves as input for the OZ equation~(\ref{eq46b}).

The fact that the OZ equation relates $c^{\circ}_{nn'}(\Omega,\Omega')$ and 
$h^{\circ}_{nn'}(\Omega,\Omega')$ (for $n \neq n'$) to each other but
the PY approximation involves $c_{nn'}(\Omega,\Omega')$ and 
$h_{nn'}(\Omega,\Omega')$ requires some discussion. In
App.~\ref{appI} we prove that $h_{nn'}(\Omega,\Omega')$ is
uniquely determined by $h^{\circ}_{nn'}(\Omega,\Omega')$ (cf.
Eq.~(\ref{eqA4})). Therefore, having determined
$h^{\circ}_{nn'}(\Omega,\Omega')$ from the OZ equation,
Eq.~(\ref{eqA4}) yields $h_{nn'}(\Omega,\Omega')$ ($n \neq n'$) and
this in turn yields $c^{\circ}_{nn'}(\Omega,\Omega')$ by use of the PY
approximation (\ref{eq47}). In order to calculate
$G_{nn',\lambda\lambda'}$ or $G_{\lambda\lambda'}({\bf q})$ one also
needs the self part of the direct correlation function. Taking $n =
n'$ in Eq.~(\ref{eq46a}) yields the matrix equation
\begin{equation}
\label{eq51}
{\bf c}^{\circ}_{nn} = -\frac{1}{4\pi} \sum\limits_{n' \neq n}
{\bf c}^{\circ}_{nn'} \, {\bf D} \, {\bf h}^{\circ}_{n'n}\,,
\end{equation}
i.e. the self part $c^{\circ}_{nn}(\Omega,\Omega')$ is determined
by the distinct parts of $c^{\circ}_{nn'}(\Omega,\Omega')$ and 
$h^{\circ}_{nn'}(\Omega,\Omega')$, only. This feature exhibits a further
difference to the OZ equation and PY approximation for liquids. If $K
\subset S^{2}$, additional new features occur which will be discussed
in the next subsection and App.~\ref{appV}. 
\subsection{\label{secIIIb}Molecular crystals characterized by $K \subset S^{2}$}
As already stressed in Sec.~\ref{secIIa}, there may exist a non-empty
subset $\bar{K}$ of $S^{2}$ on which $\rho^{(1)}(\Omega)$ and 
$G_{nn'}(\Omega,\Omega')$ vanish. This case only occurs for hard
particles, if their size exceeds a certain limit.
Accordingly, any function
$f(\Omega)$ or $F_{nn'}(\Omega,\Omega')$ must be restricted to
$\Omega,\Omega' \in K$, denoted by $f^{K}(\Omega)$ and 
$F^{K}_{nn'}(\Omega,\Omega')$, respectively. It is easy to prove that
Eqs.~(\ref{eq11}) read now for all $nn'$:
\begin{equation}
\label{eq52}
\int\limits_{K} G^{K}_{nn'}(\Omega,\Omega') \, d\Omega =
\int\limits_{K} G^{K}_{nn'}(\Omega,\Omega') \, d\Omega' = 0\,.
\end{equation}
This suggests to introduce for each function
$F^{K}_{nn'}(\Omega,\Omega')$
\begin{equation}
\label{eq53}
F^{K,\circ}_{nn'}(\Omega,\Omega') = (R^{K} F^{K}_{nn'} R^{K})(\Omega,\Omega')
\end{equation}
with the projector
\begin{subequations}
\label{eq54}
\begin{equation}
\label{eq54a}
R^{K}(\Omega,\Omega') =
\left\{
\begin{array}{c}
\delta (\Omega|\Omega') - \frac{1}{|K|}, \quad \Omega,\Omega' \in K \\
0\,, \text{     otherwise}
\end{array}
\right . 
\end{equation}
having the $\lambda$-transform
\begin{align}
\label{eq54b}
R^{K}_{\lambda\lambda'}  =  &i^{l'-l}\int\limits_{K} 
 Y_{\lambda}^{*}(\Omega) \,Y_{\lambda'}(\Omega) \, d\Omega \nonumber\\
&  -i^{l'-l} \frac{1}{|K|} \int\limits_{K} 
 Y_{\lambda}^{*}(\Omega) \,d\Omega 
\int\limits_{K} Y_{\lambda'}(\Omega') \,d\Omega' \,,
\end{align}
\end{subequations}
which are generalizations of Eqs.~(\ref{eq31}) and~(\ref{eq32}). Note
that $|K|$ is the total space angle covered
by the area $K \subset S^{2}$, and that in Eqs.~(\ref{eq20}),~(\ref{eq21}), 
$Y_{\lambda}(\Omega)$ and $Y_{\lambda'}(\Omega')$ have to be restricted to
$K$. If $K \subset S^{2}$, $\{Y_{lm}(\Omega)\}$ is no longer an orthonormal
set. The latter implies the very important fact that if $K \subset
S^{2}$ there exists no angular function $F^{K}(\Omega,\Omega')$ for
which the associated matrix ${\bf F}^{K}_{nn'}$ is the unity
matrix. Therefore, inversion of $F^{K,\circ}_{nn'}(\Omega,\Omega')$
and of ${\bf F}^{K,\circ}_{nn'}$ must be performed with respect to
$R^{K}(\Omega,\Omega')$ and ${\bf R}^{K}$, respectively:
\begin{align}
\label{eq55}
\sum\limits_{n''} \int\limits_{K} &
\left(\left(F^{K,\circ}\right)^{-1}\right)^
{K,\circ}_{nn''}(\Omega,\Omega'') \,F^{K,\circ}_{n''n'}
(\Omega'',\Omega') \,d\Omega'' \nonumber \\
& = \delta_{nn'} R^{K} (\Omega,\Omega')
\end{align}
and similar for the $\lambda$-transforms. Taking these modifications
into account one can follow each step done in
subsection~\ref{secIIIa}. Finally one obtains the OZ equation for the case $K
\subset S^{2}$ which follows from Eq.~(\ref{eq46a}) by replacement 
of $S^{2}$ and $\circ$ by $K$ and $K,\circ$, respectively. The factor
$\frac{1}{4\pi}$, however, remains. The same replacement has to be
done for all functions occuring in the PY approximation~(\ref{eq48}),~(\ref{eq50}).

So far the OZ equation and the PY approximation for $K \subset S^{2}$
are directly related to the corresponding equations for $K = S^{2}$ by
the replacements mentioned above. A new feature is related to solving 
the OZ equation 
\begin{equation}
\label{eq56}
{\bf h}^{K,\circ}({\bf q}) = {\bf c}^{K,\circ}({\bf q}) + 
\frac{1}{4\pi} {\bf c}^{K,\circ}({\bf q})\,{\bf D}\,{\bf
h}^{K,\circ}({\bf q}) \,, 
\end{equation}
for ${\bf h}^{K,\circ}({\bf q})$, which is needed for the
self-consistent solution. In the case $K = S^{2}$, one performes a
simple matrix inversion. For $K \subset S^{2}$, this can be done in
the following way. First, rewrite Eq.~(\ref{eq56}) as
\begin{align}
\label{eq57}
\left({\bf R}^K - \frac{1}{4\pi} {\bf c}^{K,\circ}({\bf q})\,{\bf
D}\right) {\bf h}^{K,\circ}({\bf q}) = {\bf c}^{K,\circ}({\bf q}) \,. 
\end{align}
Note that we have to use ${\bf R}^K {\bf h}^{K,\circ}({\bf q}) =  {\bf
h}^{K,\circ}({\bf q})$, since there exists no function
$F^{K}(\Omega,\Omega')$ with $F^{K}_{\lambda\lambda'} = \delta_{\lambda\lambda'}$. 
The first rows and columns of the matrices ${\bf R}^K$ and
${\bf R}^K - \frac{1}{4\pi}{\bf c}^{K,\circ}({\bf q})\,{\bf D}$
vanish. The inverse of the latter matrix with respect to ${\bf R}^K$
is given by ${\bf R}^K + \frac{1}{4\pi}{\bf h}^{K,\circ}({\bf
q})\,{\bf D}$, due to Eq.~(\ref{eq56}). Then it follows 
\begin{subequations}
\label{eq58}
\begin{align}
\label{eq58a}
{\bf h}^{K,\circ}({\bf q}) = \left(\left({\bf R}^K - \frac{1}{4\pi}
{\bf c}^{K,\circ}({\bf q})\,{\bf D}\right)^{-1}_{{\bf R}^K}\right)^{K,\circ}
{\bf c}^{K,\circ}({\bf q}) \,,
\end{align}
where the lower index ${\bf R}^{K}$ indicates inversion with respect to ${\bf R}^K$.
In the next lines, ${\bf 1}$ means the full unity matrix or the unity
matrix for $l,l' > 0$. It is interesting
that the matrix ${\bf 1} - \frac{1}{4\pi} 
{\bf c}^{K,\circ}({\bf q})\,{\bf D}$ still has an inverse with respect
to the unity matrix ${\bf 1}$. This inverse is ${\bf 1} +
\frac{1}{4\pi} {\bf h}^{K,\circ}({\bf q})\,{\bf D}$, according to Eq.~(\ref{eq56}).
Consequently, an equivalent form to~(\ref{eq58a}) is
\begin{align}
\label{eq58b}
{\bf h}^{K,\circ}({\bf q}) = \left({\bf 1} - \frac{1}{4\pi}
{\bf c}^{K,\circ}({\bf q})\,{\bf D}\right)^{-1}_{{\bf 1}}
{\bf c}^{K,\circ}({\bf q}) \,,
\end{align}
\end{subequations}
Indices $K$ or $\circ$ for the inverse on the rhs of
Eq.~({\ref{eq58b}) have been left out, since it describes no angular 
function like the inverse in~({\ref{eq58a}). It is easy to prove that 
the matrices ${\bf h}^{K,\circ}({\bf q})$ calculated due to 
Eqs.~(\ref{eq58}) are hermitian, if ${\bf D}$ and 
${\bf c}^{K,\circ}({\bf q})$ are.

Because of these complications for $K \subset S^{2}$, on one side
one must solve the OZ equation using PY theory on the restricted
angular space, i.e $\Omega,\Omega' \in K$. But on the other side all 
physical quantities entering the OZ and PY equations as input are 
continuous under variation of the pair potential
$V_{nn'}(\Omega,\Omega')$. Even though the physical part $K$ of
$S^{2}$ may experience a jump from $K = S^{2}$ to $K \subset S^{2}$, 
the physically relevant
structure factors~(\ref{eq40}),~(\ref{eq41}) should behave
continuously, too, if we work on $S^{2}$, assuming finite interaction $V < \infty$
whenever in reality we have hard core overlap, and then turn on the hard cores by
taking the limit $V \longrightarrow \infty$. 
In App.~\ref{appV} it will be shown that if
hard core interactions lead to $K \subset S^{2}$, one can indeed
determine solutions $h^{\infty}_{nn'}(\Omega,\Omega')$ and
$c^{\infty}_{nn'}(\Omega,\Omega')$ of the OZ/PY equations for    
$\Omega,\Omega' \in S^{2}$ and $V = \infty$,
which lead to the same structure 
factors as the functions $h^{K}_{nn'}(\Omega,\Omega')$ and 
$c^{K}_{nn'}(\Omega,\Omega')$, introduced above. Consequently, 
for practical purposes it is much easier to solve the OZ and PY
equations defined for $\Omega,\Omega' \in S^{2}$ in Sec.~\ref{secIIIa} 
even for $K \subset S^{2}$ to calculate the strucure factors
$S_{\lambda\lambda'}({\bf q})$. 
\section{\label{secIV}Results for hard ellipsoids on a sc-lattice}
The self consistent solution of the OZ equation and the PY
approximation has been done numerically. In order to check the quality
of these solutions we have performed MC simulations, which also allow
to determine the phase boundary between orientationally ordered
and disordered phases. We stress that the investigation of the phase
transition has not been our major motivation. Therefore we have not
attempted to verify the phase boundary by more complicated MC algorithms. 
Before we describe the numerical
solution of the OZ/PY equations in Sec.~\ref{secIVb}, let us
present some details of the MC simulations in the first subsection. 
The results from both approaches will be discussed in Sec.~\ref{secIVc}.

For both the numerical solution of the OZ equation using the
PY approximation and the MC studies the overlap criterion of
Vieillard-Baron~\cite{VB1972} for hard ellipsoids of revolution is
used. 

The numerical studies are done in terms of matrix elements 
for all $\lambda$ up to a certain limit $l_{\text{max}}$, 
refering to all correlation function matrices up to the $f$-function matrices, 
for which the maximum $l$-value is $2\,l_{\text{max}}$. Since the ellipsoids 
have inversion symmetry and are assumed to be fixed with their centers
of mass on the lattice, the matrices ${\bf G}_{nn'}$, ${\bf h}_{nn'}$, ${\bf c}_{nn'}$
and ${\bf f}_{nn'}$, $n \neq n'$, have non-vanishing entries for 
$l$ and $l'$ even only, whereas the matrix ${\bf D}$, and therefore also ${\bf G}_{nn}$, has
also non-vanishing entries for $l$ and $l'$ odd. Despite of these ($ll'$
odd)--entries, the OZ equation~(\ref{eq46b}) reduces to an equation
for the blocks $l,l' \ge 2$ and even.
\subsection{\label{secIVa}MC simulations}
The MC simulations are performed for systems of $16 \times
16 \times 16$ simple cubic lattice sites and periodic boundary conditions. 
For systems having very long-ranged correlations along a certain lattice
direction, test runs with $32 \times 32 \times 32$ particles are done, 
but the results are only slightly different and not shown in this
work. 

The only possible move is a rotation for a randomly chosen
particle perpendicular to its orientation axis ${\bf
u}_{n}$ by an angle $0 \le \theta \le \theta_{\text{max}} \le \pi$,
where $\cos \theta$ is at random, and a subsequent rotation with respect 
to its original orientation by a random angle $0 \le \phi \le 2 \pi$. 
If this move leads not to an overlap of ellipsoids, it is accepted,
otherwise rejected, in which case the next move is not tried for the
same particle, but for a new randomly chosen one.

As starting configuration for each MC run parallel ellipsoids are
chosen. Each particle is moved on an average of 1000 times with 
$\theta_{\text{max}} = \frac{\pi}{2}$ to get a rapid convergence 
to the \textit{disordered} phase having a cubic symmetry, 
if possible. This process can be followed measuring the
biggest eigenvalues $\lambda_{+}$ of the Saupe tensor~\cite{ST1964} for 
the instantaneous configurations at certain time values and also for 
subsystems of appropriate size. In the disordered, cubic phase 
the time average of $\lambda_{+}$ 
must decrease with the system size, since $\lambda_{+}$ is 
connected to the canonical averages $\langle Y_{2m} \rangle$, 
which must vanish due to Sec.~\ref{secIIb}. Next, each particle 
is moved on an average of further 1000 times, after each 10 times adjusting
$\theta_{\text{max}}$ by $\pm 0.02$ to get a acceptance rate of 
$\approx 25 \%$. Then, the system is equilibrated doing an average of $5\times
10^{4}$ moves per particle, which is equivalent to about $2 \times
10^{5}$ MC steps. After this, an average of further $10^{5}$ moves is 
done for each particle, after each $10$ moves storing $\lambda_{+}$ 
to calculate the correlation time $\tau$ of
\begin{align}
\Lambda(t) = \frac{\langle \lambda_{+}(t) \lambda_{+} \rangle - 
\langle \lambda_{+}\rangle \langle \lambda_{+} \rangle}
{\langle \lambda_{+} \lambda_{+} \rangle - 
\langle \lambda_{+}\rangle \langle \lambda_{+} \rangle}
\end{align}

In the production run, every $\tau \times 10$ moves per particle
the current configuration is used to measure the values
$\rho^{(1)}_{\lambda} \equiv i^{-l} \,\langle Y^{*}_{\lambda} \rangle$ 
($l = 2,4,6,8$; $m = 0, \ldots, l$) and
$\rho^{(2)}_{xyz,\lambda\lambda'} \equiv
\rho^{(2)}_{nn',\lambda\lambda'} 
\equiv \rho^{(2)}_{{\bf x}_{nn'},\lambda\lambda'}$ 
($0 \le x,y,z < 16;\, l = 2,4;\, l' \ge l; \, 0 \le m \le
l;\, -m \le m' \le m \text{ if }l =l';\, -l' \le m' 
\le l'\text{ if } l \neq l'$). See Eqs.~(\ref{eq20a}),~(\ref{eq21a})
for the definitions of $\rho^{(1)}_{\lambda}$ and
$\rho^{(2)}_{xyz,\lambda\lambda'}$. As exit condition for the
simulations, a standard deviation of $10^{-4}$ or less 
for the values $\rho^{(2)}_{xyz,\lambda\lambda'}$ has been
chosen.

After the production run,
a more serious evidence for the cubic symmetry than the behaviour of 
$\lambda_{+}$ is to check whether the measured data $\langle
Y_{\lambda} \rangle$, have about the proportions given in
Eqs.~(\ref{eq25}) and if the other values really ``vanish''.
Then, the matrix ${\bf d}$ (see Eq.~(\ref{eqB2b})) is calculated 
by Clebsch-Gordan coupling~\cite{GG1984} from
the measured data and averaged over all cubic symmetry operations. After 
this, ${\bf d}$ is composed by the best estimates for the values 
$\langle \hat{Y}_{41} \rangle$, $\langle \hat{Y}_{61} \rangle$ and 
$\langle \hat{Y}_{81} \rangle$, and its first row is used to extract these values.
In the last step of evaluation, the above missing matrix elements 
$\rho^{(2)}_{xyz,\lambda\lambda'}$
are calculated by symmetry. These are also subjected to all
cubic symmetry operations, each referring a to point symmetry operation
on the complete system, and averaged. The third test for the cubic 
symmetry is then to check if these averaged correlations match the old ones. 

The resulting phase diagrams for prolate and oblate ellipsoids are
shown in Fig.~\ref{fig2}. The thin solid lines characterize the
closest packing of parallel ellipsoids. They represent upper bounds
for the phase boundaries for transitions to ordered phases with
\textit{aligned} ellipsoids. 
Whether there exist more complex, ordered phases, commensurate or 
incommensurate ones with even larger volume fraction than
on the thin solid lines is not
known. An interesting feature of these lines can be
observed. For prolate and oblate ellipsoids there are characteristic
pairs $(a,b)$ at which cusps occur, indicating a maximum volume
fraction. The light grey areas represent the transition region from an
orientationally disordered to an ordered phase. The latter is not
necessarily a phase of aligned ellipsoids.
Transitions have been observed from the ordered to the
disordered phases and also vice versa for systems with ellipsoids
of small enough maximum linear dimension, having only few interaction
partners. The hysteresis is small, indicating either a continuous or a
weakly first order phase transition. The medium grey areas refer to
the OZ/PY solution and are explained in the next Sec.~\ref{secIVb},
while the dark grey areas indicate overlap of the light and medium grey
ones.
\begin{figure}[t]
\includegraphics[bb = 50 0 580 830 ,width=9.5cm]{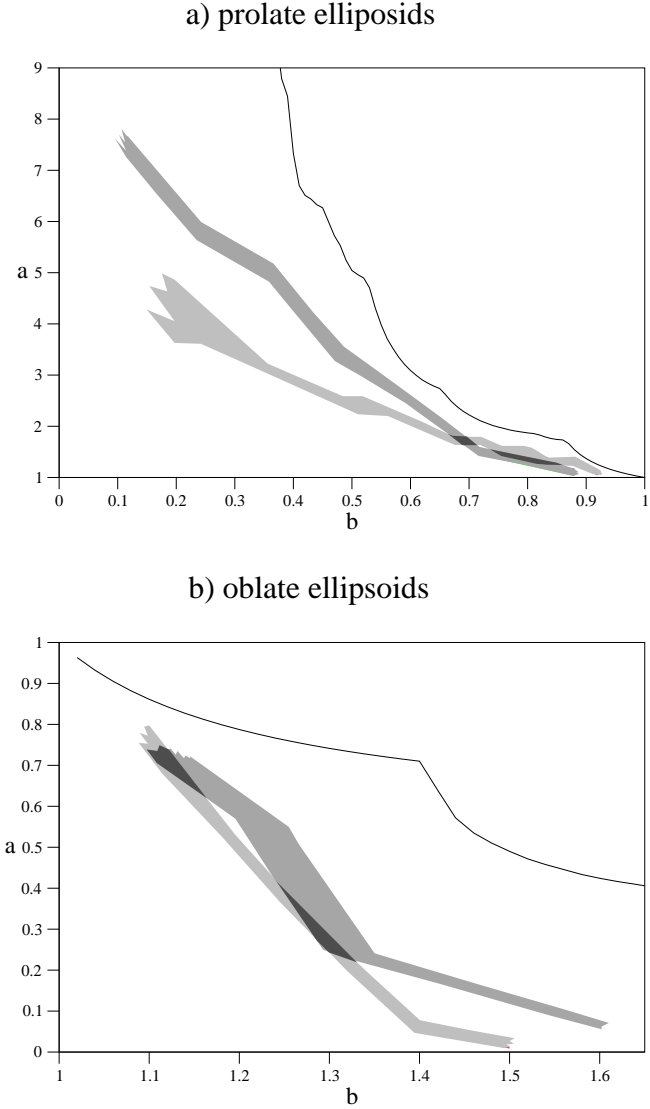}
\caption{\label{fig2}Phase diagrams for a) prolate and b) oblate
hard ellipsoids of revolution on a simple cubic lattice. 
Solid lines refer to the closest packing for parallel ellipsoids.
Within the light grey areas an order-disorder
phase transition occurs, while the medium grey areas indicate where
the numerical solution of the OZ/PY equations starts to diverge.
For a more detailed discussion see text of 
Secs.~\ref{secIVa} and~\ref{secIVb}.}
\end{figure}
\subsection{\label{secIVb}Numerical solution of the OZ equation using the
PY approximation}
In App.~\ref{appV}, we show that the correct static structure factors 
$S_{\lambda\lambda'}({\bf q})$, even for $K \subset S^{2}$, can be
obtained by solving the OZ/PY equations for $\Omega,\Omega' \in
S^{2}$. We have chosen this option since it avoids the use of the
projectors $R^{K}$ and $P^{K}$ (cf. Sec.~\ref{secIIIb} and
App.~\ref{appII}). We also solved both equations for $\Omega,\Omega' \in
K$ as described in Sec.~\ref{secIIIb}, but it turned out that the
results are worse. The reason is probably the fact that the cut-off
$l_{max}$, at which the matrix equations are truncated, leads
to a rather crude approximation for $R^{K}$ and $P^{K}$. For example,
some of the truncated projectors are significantly non-idempotent. 

The numerical solution of the OZ/PY equations is performed by
the iterative procedure described below usually for lattices of 
size $32 \times 32 \times 32$, periodic boundary conditions and $l_{\text{max}} = 4$. 
The correlation function matrices ${\bf G}_{xyz} \equiv 
{\bf G}_{{\bf x}_{nn'}} \equiv {\bf G}_{nn'}$ 
have, by symmetry and periodicity, only to be 
tabulated for $0 \le x \le y \le z \le 16$, and the same for the 
discrete Fourier transformed matrices ${\bf S}({\bf q})$. 

For some systems characterized by $(a,b)$, we have solved the OZ/PY 
equations additionally for $l_{\text{max}} = 2$ and $l_{\text{max}}=
6$. All the correlators we have investigated explicitely remain
qualitatively unchanged, while the correlation lengths of these 
correlators become larger for increasing $l_{max}$. For some of the
systems showing no convergence of the iteration scheme 
because of diverging correlation lengths
for $l_{\text{max}} = 4$ and $32 \times 32 \times 32$ lattice sites, 
system sizes up to $128 \times 128 \times 128$ have been
used, but for none of these bigger system sizes convergence could be
achieved. 

After each numerical inversion or Fourier transform the 
symmetries are checked to be within certain error margins, and then 
each matrix is subjected to every symmetry
operation under which it must be invariant. From all these 
symmetry-transformed matrices the average is calculated to rule 
out rounding errors to become too large.

Before starting the iterative procedure, the input quantities are
calculated:

 (a) For each pair $(a,b)$ of ellipsoid axes, the $f$-function 
matrix elements $f^{r}_{ll'm}$ in the $r$-frame~\cite{GG1984} according to 
Eq.~(\ref{eqC5}) have to be calculated for all even 
$l \le l'\le 2\,l_{\text{max}}$ and $0 \le m \le
\text{max}(l,l')$. By Eqs.~(\ref{eqC1}), the
remaining elements are obtained. Then, these are transformed back to
the laboratory frame. For the numerical integration a $400 \times 200 \times
200$--grid has been used.

(b) The values $\langle \hat{Y}_{41}\rangle$, $\langle
\hat{Y}_{61}\rangle$ and $\langle \hat{Y}_{81}\rangle$ are taken from
the MC simulation for the same pair $(a,b)$ or an appropriate
extrapolation beyond the MC cubic phase boundary, where needed. Then,
Eqs.~(\ref{eq25}) yield the canonical averages of the spherical
harmonics, which are coupled to give the matrices ${\bf d}$ and ${\bf
D}$, due to Eqs.~(\ref{eqB2b}) and~(\ref{eqB2c}). 

The iterative procedure is now as follows:

(a) As input for the OZ equation in the $n$-th iteration step, 
we use the matrices ${\bf c}^{(n)'}_{xyz}$, which will be specified below. 
According to Eqs.~(\ref{eq31b}) and ~(\ref{eq32b}),
the matrices ${\bf c}^{\circ,(n)'}_{xyz}$ are
obtained by just cancelling the first columns and rows. 
The matrix elements $c^{\circ,(n)'}_{xyz,\lambda\lambda'}$ are then
calculated for $0 \le x,y,z < 32$ and subsequently Fourier 
transformed to yield $c^{\circ,(n)'}_{pqr,\lambda\lambda'}$. 
As very first input we choose 
${\bf c}^{(1)'}_{xyz=0} = {\bf 0}$, ${\bf c}^{(1)'}_{xyz} = {\bf
f}_{xyz}$ otherwise.

(b) The matrices ${\bf h}^{\circ,(n)'}_{pqr}$ in the $n$-th iteration step
are calculated from
\begin{equation}
{\bf h}^{\circ,(n)'}_{pqr} = \left( \left({\bf 1} - \frac{1}{4\pi} {\bf
c}^{\circ,(n)'}_{pqr} {\bf D} \right)^{-1}_{\bf 1} \right)^{\circ} {\bf
c}^{\circ,(n)'}_{pqr}\,,\quad n \ge 1\,.
\end{equation}
They generally will not fulfill the constraint 
${\bf h}^{\circ,(n)'}_{xyz=0} = {\bf 0}$. Therefore, we define matrices
${\bf h}^{\circ,(n)}_{pqr} = {\bf h}^{\circ,(n)'}_{pqr}-{\bf
h}^{\circ,(n)'}_{xyz=0}$, which fulfill the constraint. 
Substituting ${\bf h}^{\circ,(n)}_{pqr}$ into the OZ equation
yields matrices ${\bf c}^{\circ,(n)''}_{pqr}$, from which
${\bf c}^{\circ,(n)''}_{xyz=0}$ is obtained. Then the final 
$n$-th approximation of the direct correlation function is chosen to be
\begin{equation}
\label{eq61}
{\bf c}^{(n)}_{xyz}= 
\left\{
\begin{array}{cc}
\alpha \, {\bf c}^{(n)''}_{xyz} + (1-\alpha)\,
{\bf c}^{(n)'}_{xyz} \,, & xyz=0\,, \\
{\bf c}^{(n)'}_{xyz}\,, & \text{otherwise}\,.
\end{array}
\right .
\end{equation}
Note that the self part of the direct correlation function can definitely
be determined in the form $\circ$ only (see Eq.~(\ref{eq51})).
In Eq.~(\ref{eq61}), $0 < \alpha < 1$ is a mixing parameter.

(c) ${\bf h}^{\circ,(n)}_{xyz}$ is obtained from ${\bf
h}^{\circ,(n)}_{pqr}$ by Fourier transform and Eq.~(\ref{eqB7}) yields ${\bf
h}^{(n)}_{xyz}$ for $xyz \neq 0$. Then, ${\bf
g}^{(n)}_{xyz}$ can be calculated following App.~\ref{appII}. Using
Eq.~(\ref{eq50}), ${\bf g}^{(n)}_{xyz}$ and ${\bf c}^{(n)}_{xyz}$
for $xyz \neq 0$, one obtains new direct correlation function matrices, 
which are mixed to a fraction of $\alpha$ with the matrices ${\bf
c}^{(n)}_{xyz}$ to give ${\bf c}^{(n+1)'}_{xyz}$, 
while  ${\bf c}^{(n+1)'}_{xyz=0} = {\bf c}^{(n)}_{xyz=0}$ is chosen. 

This prodedure is repeated until a fix point for the matrices
has been reached. Typically, $\alpha = 0.1$ is 
chosen to avoid divergence. Convergence is assumed if all elements of 
${\bf h}^{\circ,(n)'}_{xyz=0}$ have submerged a certain treshold, 
which is chosen to be $10^{-13}$ times the maximum absolute value of 
any matrix element $h^{\circ,(n)'}_{xyz,\lambda\lambda'}$. 
Additionally,  the average of all non-zero matrix elements of 
${\bf c}^{(n+1)}_{xyz} - {\bf c}^{(n)}_{xyz}$ must be below
$\alpha$ times a second treshold, which is calculated in the same
manner as the $h$-treshold, but by taking also the $ll'=0$
matrix elements into account. It is interesting that Eq.~(\ref{eq51})
cannot be used for the calculation of ${\bf c}^{\circ,(n)}_{xyz=0}$, since
this always leads to a divergence.

After the iteration has converged, also Eq.~(\ref{eq51})
can be checked to be true. For some systems, only a mixing 
parameter of $\alpha = 0.05$ or $\alpha = 0.02$ leads to convergence. 
For some of these critical systems the mixing is not enough 
for the direct correlation function, but also the new total
correlation function has to be accepted only to a fraction of $\alpha$.
The medium grey areas in Figs.~\ref{fig2} are the areas where the 
iterative procedure above, for a mixing paramter 
$\alpha = 0.02$ applied to both the new direct and the new 
total correlation functions, turns from convergent to divergent, or
begins to yield unphysical negative diagonal structure factors (see
the next section for explanation).

Finally, Eq.~(\ref{eq41}) is used to calculate the static structure
factors.
\subsection{\label{secIVc}Numerical results for the correlation functions}
Results obtained from the numerical solution of the OZ/PY
equations and the MC simulations are presented in
Figures~\ref{fig3}-\ref{fig13} for five different pairs of $(a,b)$,
including prolate and oblate ellipsoids. We have restricted the
illustrations of correlators in direct and reciprocal space
to the matrix elements $(l = l' =2, m = m' = 0,1,2)$, $(l = 2, l' =4,
m = m' = 0,1,2)$ and $(l = l' =4, m = m' = 0,1,2,3,4)$. 

Log-lin representations of 
the direct space orientational correlators $G_{xyz,\lambda\lambda'}$ are
shown along lattice directions of high symmetry, i.e. $(x,y,z) =
n\,(0,0,1)$, $(x,y,z) = n\,(0,1,1)$ and $(x,y,z) = n\,(1,1,1)$ for 
$n =0 ,1,\ldots, 8$ (part a) of Figs.~\ref{fig3}-\ref{fig13}). 
Along these directions, 
all $G_{xyz,\lambda\lambda'}$ are real by the symmetries~(\ref{eq27}) 
for the chosen $\lambda\lambda'$.
Note that the PY/OZ results could in principle 
be displayed up to $n = 16$ and that a step $\Delta n =1$ 
corresponds to different lengths in
direct space, namely $1$, $\sqrt{2}$ and $\sqrt{3}$ for the different
lattice directions. In these illustrations, the values $n$
are chosen as abscissas. For each $m = m'$ and each lattice 
direction, a separate picture is provided and a logarithmic plotting 
has been chosen for positive and negative values 
of $G_{xyz,\lambda\lambda'}$ separately, i.e. the negative values are
presented as $-\log\, |G_{xyz,\lambda\lambda'}|$. This plotting shows that the 
direct space correlations decay exponentially in most cases. The respective values
of $xyz,\lambda\lambda'$ are included, too. Note that
the scatter of the MC data for some $\lambda\lambda'$ and 
large enough values of $n$ is due to the error margins.

Similarly to direct space, only correlators $S_{\lambda\lambda'}({\bf q})$
along highly symmetric reciprocal lattice directions  are
displayed, i.e. $(p,q,r) = \xi
(0,0,\pi)$, $(p,q,r) = \xi (0,\pi,\pi)$ and $(p,q,r) =
\xi(\pi,\pi,\pi)$ for $\xi =0 \ldots 1$, which are the
correlators from the Brillouin zone center to its edge in the
respective direction (part b) of
Figs.~\ref{fig3}-\ref{fig13}). In these figures, $\xi$ is chosen as
abscissa and the curves for all three reciprocal lattice directions
have been put into one illustration for each pair $\lambda\lambda'$,
which is included in each picture. The curves are distinguished by
the symbols $\Delta$ ($(0,0,\pi)$-direction), $\Sigma$
($(0,\pi,\pi)$-direction) and $\Lambda$ ($(\pi,\pi,\pi)$-direction).
All $S_{\lambda\lambda'}({\bf q})$ which are shown are real by the
symmetries~(\ref{eq28}). Additionally, by Eq.~(\ref{eq40}), the diagonal elements 
$S_{2m2m}({\bf q})$ and $S_{4m4m}({\bf q})$ are positive. The 
numerically determined correlators $S_{\lambda\lambda'}({\bf q})$ have been interpolated by cubic
splines with the correct boundary condition of vanishing gradients for $\xi
=0$ and $\xi =1$. Note the different scales of the illustrations for different
$\lambda\lambda'$.

\begin{figure}
\includegraphics[angle=270,width=8.5cm]
	{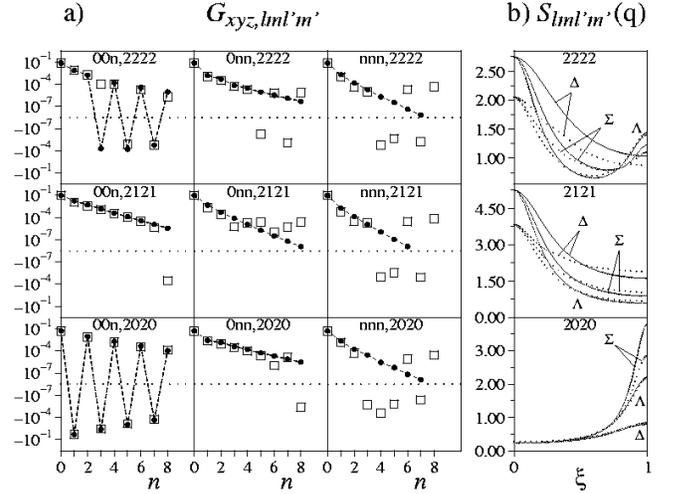}
\caption{\label{fig3}a) Log-lin-representation of the
direct space orientational correlators
$G_{xyz,2m2m}$ along highly symmetric lattice directions
(solid circles = OZ/PY results, squares = MC results; 
dashed lines are a guide to the eye),
b) orientational structure factors $S_{2m2m}({\bf q})$ along the respective reciprocal lattice
directions (solid lines = OZ/PY results, dotted lines = MC
results). These results are for oblate hard ellipsoids with axes $a = 0.4$, $b = 1.2$
on a sc lattice and $m=0,1,2$. For further explanation see text of Sec.~\ref{secIVc}.}
\end{figure}
\begin{figure}
\includegraphics[angle=270,width=8.5cm]
	{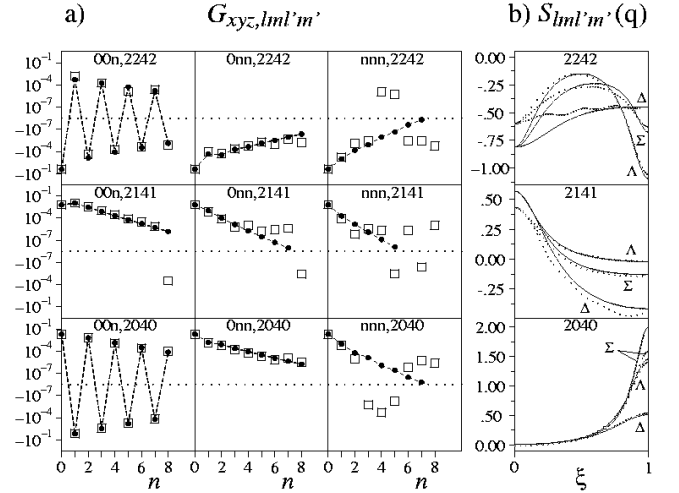}
\caption{\label{fig4}Same as Figure~\ref{fig3}, but for $l=2$, $l'=4$.}
\end{figure}
\begin{figure}
\includegraphics[angle=0,width=8.5cm]
	{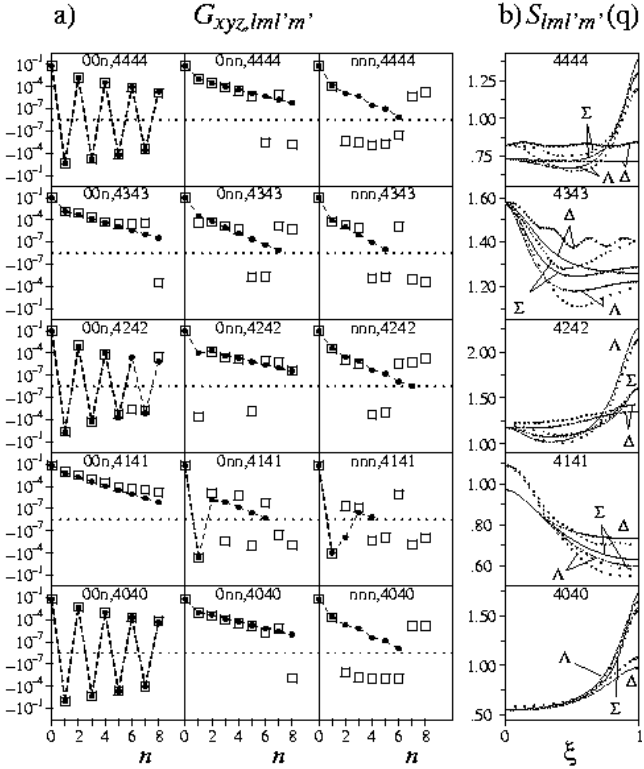}
\caption{\label{fig5}Same as Figure~\ref{fig3}, but for $l=l'=4$ and $m=0,1,2,3,4$.}
\end{figure}
Let us start by discussing the results shown in
Figs.~\ref{fig3}-\ref{fig5} for oblate hard ellipsoids with axes
$a =0.4$ and $b=1.2$, corresponding to a packing fraction of $\phi \approx 0.3$.
This system is quite close to the MC phase boundary
shown in Fig.~\ref{fig2} b), but not close enough to find a tendency to a
divergence of some of the $S_{\lambda\lambda'}({\bf q})$. Notice the almost perfect 
agreement of the OZ/PY and MC orientational correlators 
in direct space, in case where the MC results are large enough. In fact,
such an agreement appears for all investigated oblate ellipsoid
systems, for which MC results are available, except for the system with $a =0.72$, $b=1.1$. 
Relatively long ranged oscillations appear for all correlators along the
fourfold lattice direction $[0,0,1]$ having even $m =m'$.
The other correlators along the same direction decay
faster and monotonously without oscillation, and also the correlators along the other
directions decay without oscillations. Note that for 
$a =0.4$ and $b=1.2$ the ellipsoids can only interact via their nearest
neighbours, which are localized along the fourfold lattice 
directions. Therefore, it is tempting to assume that the oscillations 
are primarily related to a direct particle interaction
via nearest neighbours along a certain lattice direction. For denser
oblate and some prolate systems, 
the oscillations extend up to many lattice constants, indicating the
tendency to build up a medium range orientational order.
For the structure factors $S_{\lambda\lambda'}({\bf q})$, the
agreement of OZ/PY and MC results is satisfactory. The most
significant deviations appear for $l=l'=2$ near ${\bf q} = {\bf 0}$.
The oscillations exhibited by some of the $G_{xyz,\lambda\lambda'}$
manifest themselves in some maxima at the
Brillouin zone edge, mainly for correlators $S_{2m2m}({\bf q})$ and 
$S_{4m4m}({\bf q})$ with $m$ even, but also for the correlators 
$S_{2040}({\bf q})$. Increasing $a$ and/or $b$ for oblate systems, 
the OZ/PY results for some of these zone-edge maxima show a 
tendency to diverge, accompanied by a simultaneous divergence mainly of 
the remaining $l = l' =2$ correlators at the zone center.

\begin{figure}
\includegraphics[angle=270,width=8.5cm]
	{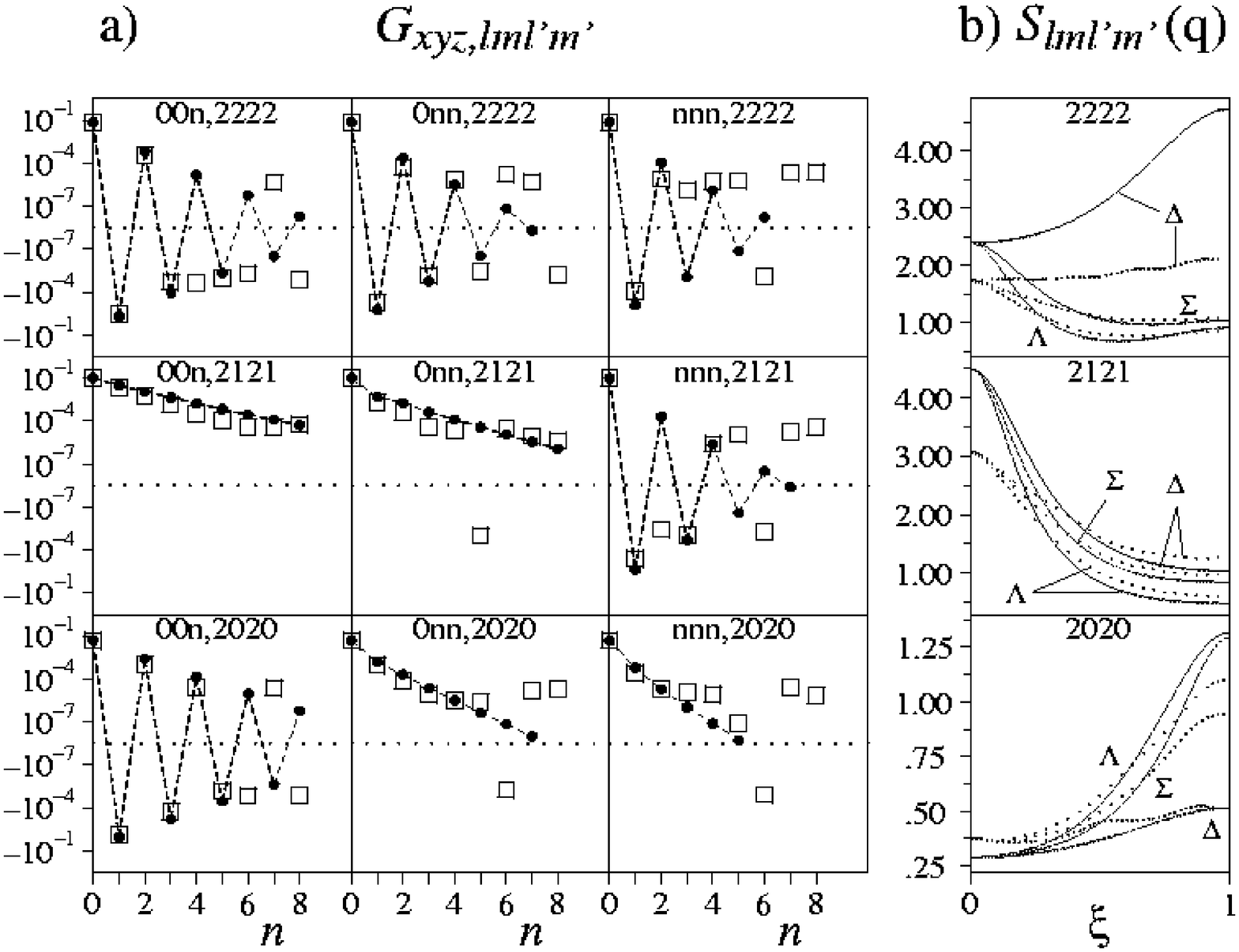}
\caption{\label{fig6}Same as Figure~\ref{fig3}, but for prolate hard
ellipsoids with axes $a = 1.6$, $b = 0.6$.}
\end{figure}
\begin{figure}
\includegraphics[angle=270,width=8.5cm]
	{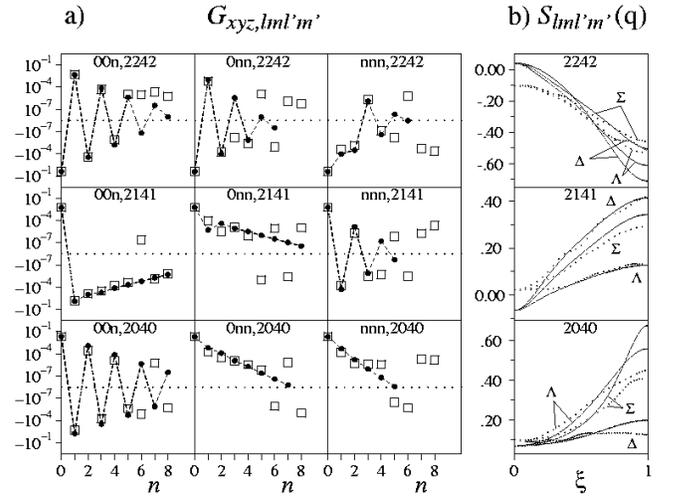}
\caption{\label{fig7}Same as Figure~\ref{fig3}, but for $l=2$, $l'=4$ 
and prolate hard ellipsoids with axes $a = 1.6$, $b = 0.6$.}
\end{figure}
Next, we have chosen the prolate ellipsoid system with axes $a =1.6$,
$b=0.6$, having a packing fraction of $\phi \approx 0.3$, for which some results are 
shown in Figs.~\ref{fig6} and~\ref{fig7}. For this system,
oscillations appear additionally along
the two other lattice directions (i.e. $[0,1,1]$ and $[1,1,1]$), which
confirm the assumption that they may be caused by direct interaction
via nearest neighbours for appropriate values of $b$ and not too large $a$. 
Again, the MC results in direct space match the analytical results
very well, though latter overestimate some of the correlation 
lengths. This is clearly seen, for example, from the decrease of the
correlators $G_{00n,2121}$ for increasing $n$, which is
different for MC and OZ/PY results. The 
much too large OZ/PY results for the correlators $S_{2222}({\bf q})$ at 
the zone boundary along the $\Delta$-direction are perhaps also due to
this overestimation. 
OZ/PY pretends an intersection of the curves for $S_{2040}({\bf q})$ 
along lattice directions $\Lambda$ and
$\Sigma$ at $\xi \approx 0.8$, which is not present in the MC results. Note that
the correlators $G_{0nn,2141}$ and $G_{nnn,2242}$ deviate significantly from the regular decrease
inherent to other correlators. 

\begin{figure}
\includegraphics[angle=270,width=8.5cm]
	{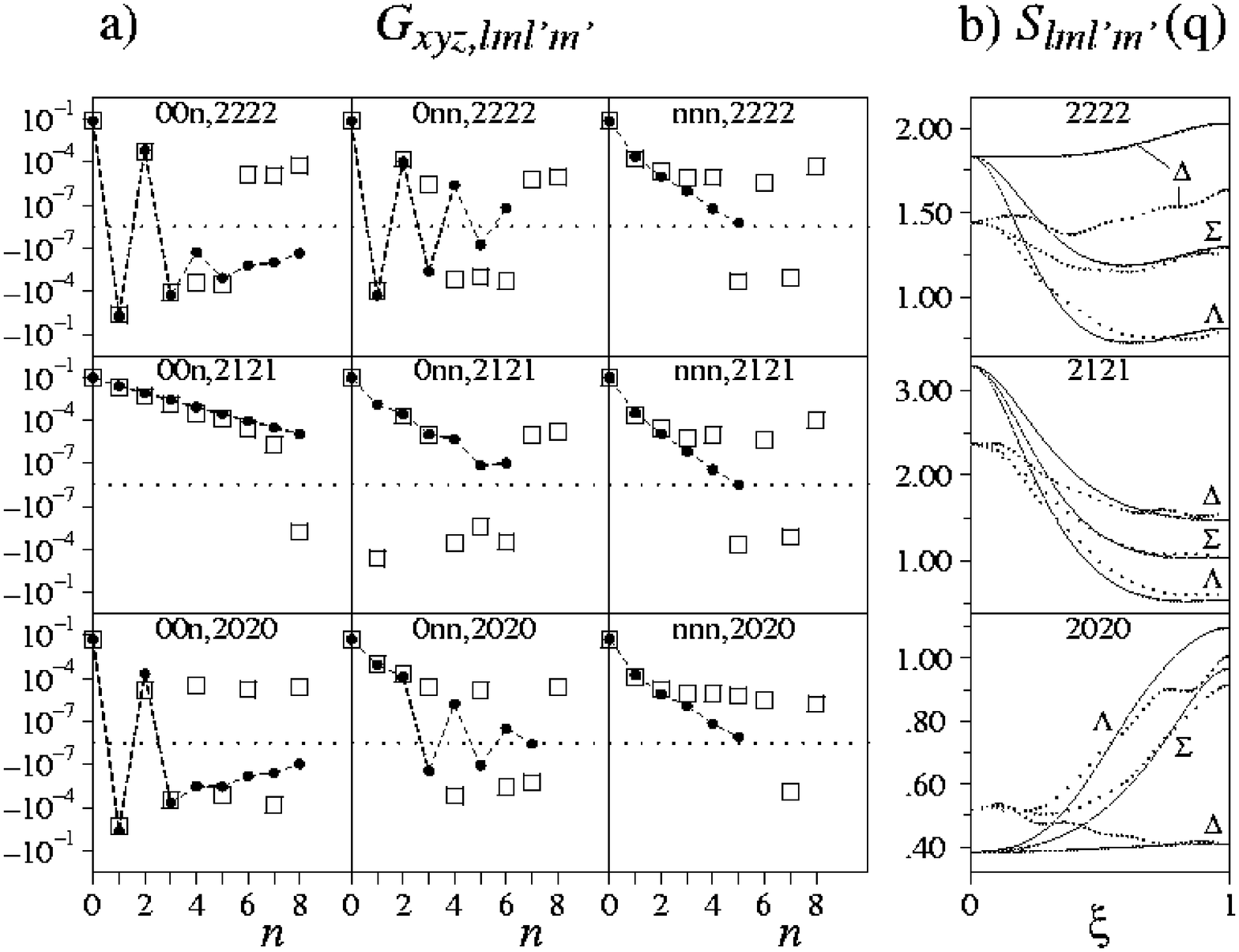}
\caption{\label{fig8}Same as Figure~\ref{fig3}, but for prolate 
hard ellipsoids with axes $a = 2.0$, $b = 0.48$.}
\end{figure}
\begin{figure}
\includegraphics[angle=270,width=8.5cm]
	{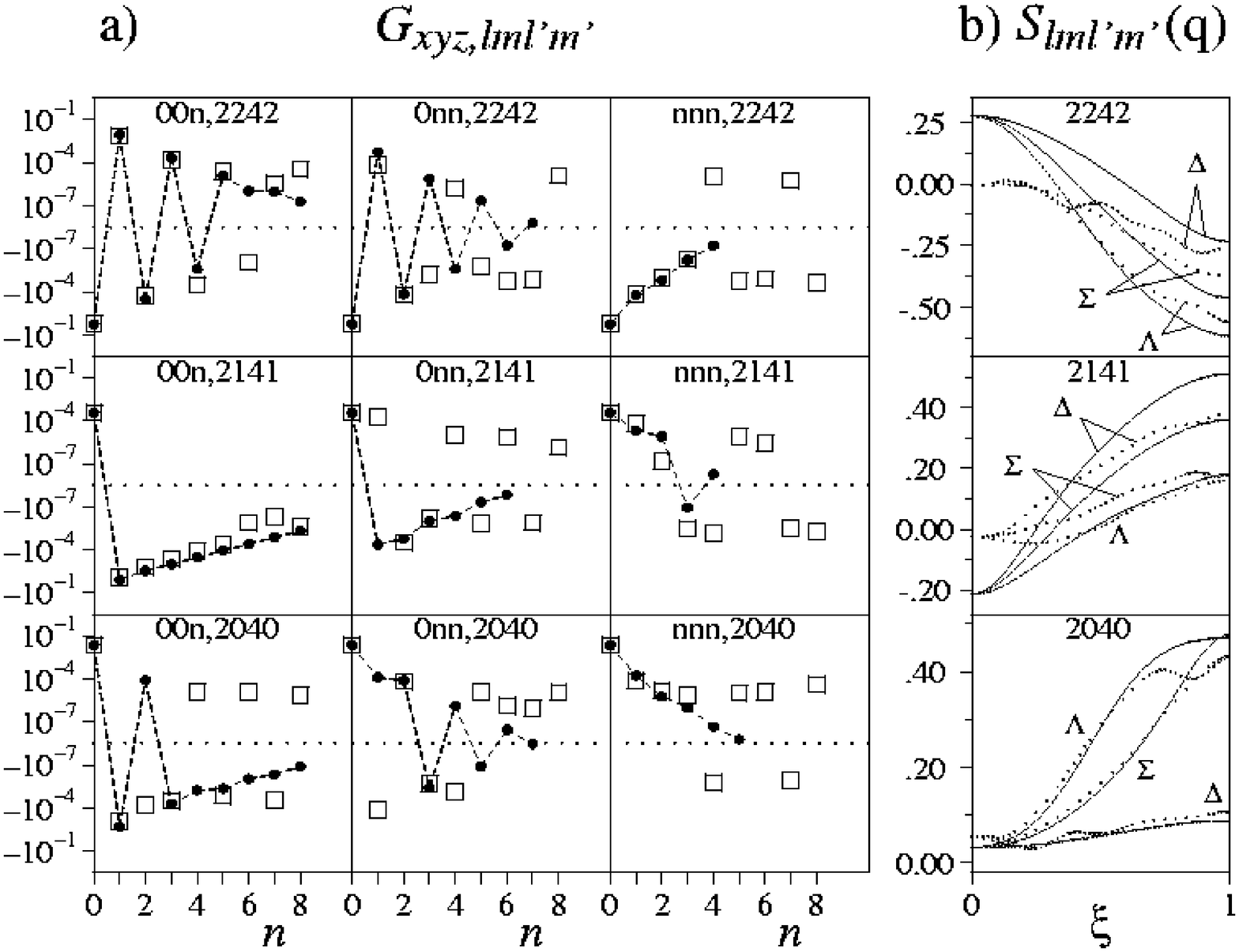}
\caption{\label{fig9}Same as Figure~\ref{fig3}, but for $l =2$, $l'
=4$ and prolate hard ellipsoids with axes $a = 2.0$, $b = 0.48$.}
\end{figure}
\begin{figure}
\includegraphics[angle=0,width=8.5cm]
	{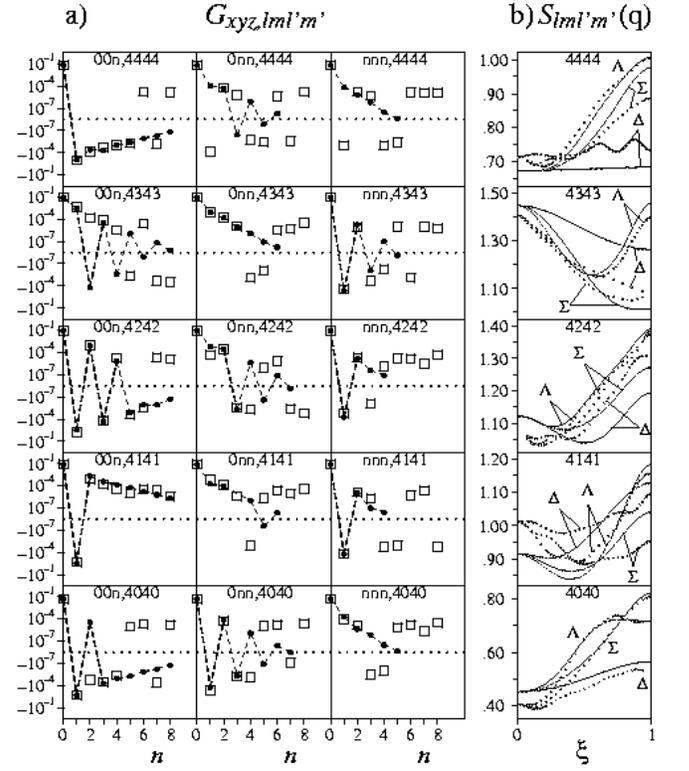}
\caption{\label{fig10}Same as Figure~\ref{fig3}, but for $l = l'=4$,
$m =0,1,2,3,4$ and prolate hard ellipsoids with axes $a = 2.0$, $b = 0.48$.}
\end{figure}
Increasing $a$ to $2.0$ and decreasing $b$ to $0.48$ leads to $\phi
\approx 0.24$ and the
results of Figs.~\ref{fig8}-\ref{fig10}. Now, only one of the
$l=l'=2$ direct space correlators still shows pure oscillating behavior ($G_{0nn,2222}$), and
also for the other values of $l,l'$ the number of
oscillating direct space correlators has
decreased. Interestingly, some correlators oscillate for small values of
$n$ and then start to decay monotonously (for example $G_{00n,2020}$,
$G_{00n,2222}$, $G_{00n,2242}$), while others oscillate from $n =2$ on 
(for example $G_{0nn,2020}$, $G_{0nn,2040}$).
Though not few of the direct space MC correlators
differ by sign from the OZ/PY correlators (for example $G_{011,2121}$), the 
results still match satisfactory. Indeed, for some of the correlators
$S_{4m4m}({\bf q})$ OZ/PY predicts crossings of the
curves for different lattice directions, which are also present in the
MC results. Coming back to the direct space orientational correlators, more and
more of them have become non-regular functions of $n$, and also
significant deviations from the almost exponential decay, which is
present in many of the direct space correlators of the other systems
discussed above, are encountered (see, e.g., $G_{0nn,2121}$). 
But the MC results for the correlators $G_{00n,4141}$, for example, 
prove that these non-regularities are not just an artefact 
of the analytical approach. Surprisingly, the OZ/PY correlator 
$G_{033,4141} \approx -2.97 \times 10^{-9}$ is to small to be
displayed in the pictures, though the surrounding correlators 
(i.e. $G_{0nn,4141}$, n = 0,1,2,4,5,6) clearly are not.

\begin{figure}
\includegraphics[angle=270,width=8.5cm]
	{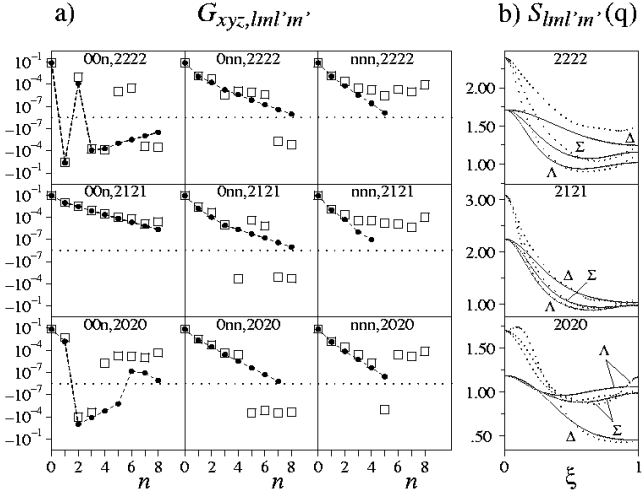}
\caption{\label{fig11}Same as Figure~\ref{fig3}, but for prolate hard
ellipsoids with axes $a = 3.6$, $b = 0.24$.}
\end{figure}
Now we pass to more and more elongated prolate ellipsoids and investigate what
happens. In the limit of this process one obtains the
hard-needle system, which is discussed, e.g., in~\cite{RLB1995}. 
Ellipsoids with axes $a = 3.6$
and $b= 0.24$ have a huge aspect ratio of $X_{0} = \frac{a}{b} = 15$.
Of course, the packing fraction of these ellipsoids on the sc
lattice is quite low, about $10 \%$, but the frustration effect of the
rigid lattice may provide completely new effects in comparison to a
liquid. As shown in Fig.~\ref{fig11}, the behavior of the correlators 
$G_{xyz,2m2m}$ is more regular in comparison to $a = 2.0$ and $b =
0.48$. Oscillations have disappeared
completely. The OZ/PY results clearly underestimate the $S_{2m2m}({\bf
q})$--correlators for small ${\bf q}$. This is also
the case for the $S_{2m2m}({\bf q})$--correlators
of all other investigated systems in the neighbourhood of $a = 3.6$, $b= 0.24$. 

\begin{figure}
\includegraphics[angle=270,width=8.5cm]
	{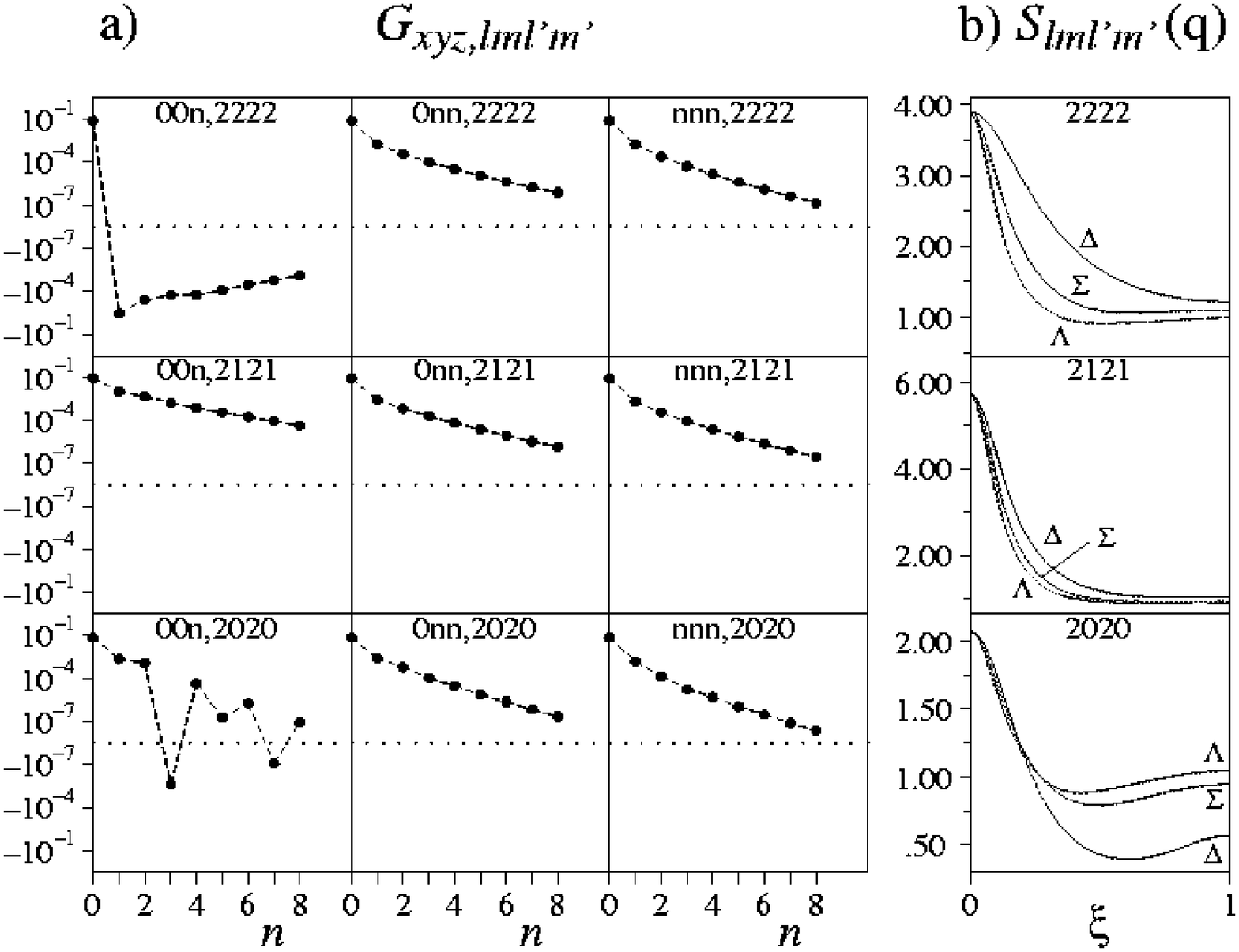}
\caption{\label{fig12}Same as Figure~\ref{fig3}, but for prolate hard
ellipsoids with axes $a = 4.8$, $b = 0.24$.}
\end{figure}
\begin{figure}
\includegraphics[angle=270,width=8.5cm]
	{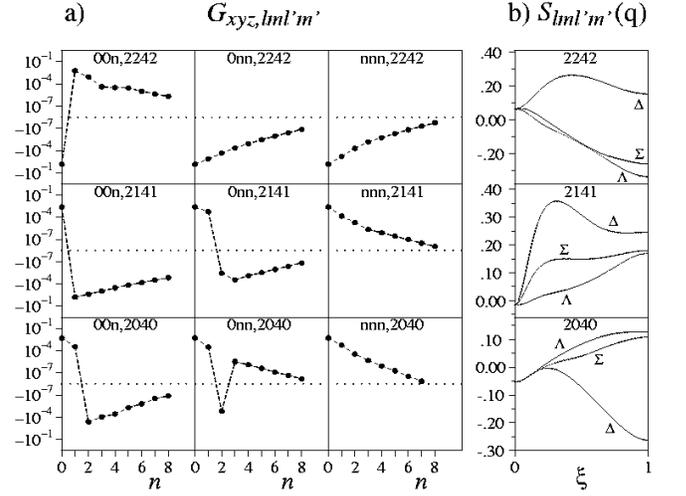}
\caption{\label{fig13}Same as Figure~\ref{fig3}, but for $l=2$, $l'=4$
and prolate hard ellipsoids with axes $a = 4.8$, $b = 0.24$.}
\end{figure}
The last system we present consists of ellipsoids  of axes $a = 4.8$
and $b= 0.24$, for which it is $X_{0} = 20$ and
$\phi \approx 16\%$. The results of Figs.~\ref{fig12} and~\ref{fig13}
belong to this pair $(a,b)$. The correlators in direct space essentially
show up monotonous decay, and the correlation lengths have clearly
increased in comparison to the elliposids of axes $a = 3.6$
and $b= 0.24$. Unfortunately, this system lies beyond the MC phase
boundary (see Fig.~\ref{fig2}a), so that no MC results are available.
Despite of the monotonous behavior of most correlators, the 
$G_{00n,2020}$--correlators, for example, still show non-regular behavior, 
which vanishes for the most part switching to $a = 5.6$ and 
$b= 0.24$ (these results are not shown here). In Fig.~\ref{fig12}, the
beginning of a divergence of the $S_{2121}({\bf q} = {\bf
0})$--correlator is seen, the corresponding value 
of the ($a = 5.6, b= 0.24$)-system being $\approx 20$. Also, what can be seen
from Fig.~\ref{fig13}, the $l=2$, $l'=4$ correlators have
got significant structure, and they seem not to diverge
like the $l=l'=2$ correlators.

\section{\label{secV}Discussion and conclusions}
Our main goal has been the study of static orientational correlation
functions for a molecular crystal in its disordered phase. For this,
we have derived the OZ equation, well-known in liquid
theory, for a rigid periodic lattice with internal orientational
degrees of freedom. As a closure relation we have adopted the
PY approximation. As pointed out, there are differences of
the present approach to that for liquids. One of them is the fact that
the OZ equation only involves the direct and total correlation
functions $c^{\circ}_{nn'}(\Omega,\Omega')$ and
$h^{\circ}_{nn'}(\Omega,\Omega')$, whereas the PY approximation
relates $c_{nn'}(\Omega,\Omega')$ and $h_{nn'}(\Omega,\Omega')$. The
functions with the superscript $\circ$ do not contain a constant part
with respect to $\Omega$ or $\Omega'$. Another
important difference is the one-particle orientational distribution function
$\rho^{(1)}(\Omega)$. In the isotropic phase of a molecular liquid it
is $\rho^{(1)}(\Omega) = \frac{1}{4\pi}$, but due to the anisotropy of
a crystal $\rho^{(1)}(\Omega)$ exhibits a nontrivial $\Omega$-dependence. In order to
solve the OZ/PY equations, one has to calculate $\rho^{(1)}(\Omega)$
separately. In our case, we have performed MC simulations. Analytical
approaches are also possible, e.g. for fixed $a$ one could perform a
kind of virial expansion for small $b$.

Despite these differences, the form of the OZ equation for a molecular
crystal is quite similar to that for molecular
liquids~\cite{GG1984}. In order to explore the applicability of the
lattice OZ equation in combination with the PY approximation, we have
solved these equations for hard ellipsoids of revolution on a simple
cubic lattice. Due to the orientational degrees of freedom, the
orientational correlators $G_{nn',\lambda\lambda'}$ in direct space or 
$S_{\lambda\lambda'}({\bf q})$ in reciprocal space, with $\lambda =
(lm)$, are tensorial quantities. Accordingly, the self consistent
numerical solution of the OZ/PY equations requires a truncation at
$l_{\text{max}}$. We mainly have chosen $l_{\text{max}} =4$. As a
result, we have found orientational correlators which have less
structure in direct and reciprocal space than for liquid systems.
Nevertheless, there are some interesting features depending
on the length $a$ and $b$ of the ellipsoid axes. For oblate ellipsoids
and prolate ones of large enough $b$, some of the direct space
orientational correlators exhibit oscillations in certain lattice
directions. Since the oscillations have period two they lead to maxima
of $S_{\lambda\lambda'}({\bf q})$ at the Brillouin zone edge for 
some $\lambda\lambda'$. Although
no long range orientational order exists, the oscillatory behavior
originates from an alternating reordering of the ellipsoids on a
finite length scale, which can extend up to many lattice constants. 

Decreasing for prolate ellipsoids $b$ and increasing $a$ leads to a
disappearance of almost all of these significant oscillations. 
In this case, the correlators
$S_{2m2m}({\bf q})$ take their absolute maxima at the Brillouin zone
center (cf. Figs.~\ref{fig11} and~\ref{fig12}), indicating
nematic-like fluctuations, while 
the same behavior is not found to the same extent for the other correlators (i.e. 
$S_{2m4m}({\bf q})$ (cf. Fig.~\ref{fig13}) and $S_{4m4m}({\bf q})$
(cf. Fig.~\ref{fig10})). The behavior of the correlators $S_{2m2m}({\bf q})$
resembles a liquid of ellipsoids with aspect ratio larger than about
two which forms a nematic phase~\cite{FMT1984}. Surprisingly,
increasing $a$ for fixed $b$ more and more the OZ/PY results for 
$S_{2m2m}({\bf q})$ lead to a divergence at ${\bf q} = {\bf
0}$, which indicates the tendency to establish a long range nematic-like order. 
This finding demonstrates that PY theory can yield the
onset of a phase transition to an ordered phase as it was already
found before for a liquid of hard ellipsoids~\cite{LL1999}. 

Some correlators for appropriately long prolate ellipsoids
show up highly irregular behavior in direct space, which is nevertheless
consistent with MC results (where available) and therefore has to be
taken serious. We suppose the frustration effect of the lattice to be
the reason for this non-regular behavior.

Comparison of the PY results with those from MC simulations shows a
satisfactory agreement. But the quality of this agreement is less good
as it is, e.g. for a liquid of hard spheres. The reason for this may
lie in the PY approximation. For a liquid its physical content has
been elucidated by Percus~\cite{P1962} (see also
ref.~\cite{HMD1990}) by use of a grand canonical ensemble. Since in a
molecular crystal the particles are fixed, it is not obvious how this
reasoning  can be used. Of course, it
might be interesting to investigate whether other closure
relations~\cite{HMD1990} or the turn to much higher $l_{\text{max}}$
can lead to an improvement.

To conclude, we have demonstrated that an extension of the OZ equation
in combination with the PY approximation to molecular crystals leads
to satisfactory results compared with MC data.
\begin{acknowledgments}
We gratefully acknowledge helpful discussions and the support in
performing the MC simulations by J. Horbach, M. M\"uller and W. Paul.
\end{acknowledgments}
\appendix
\begin{widetext}
\section{\label{appI}Calculation of $h^{K}_{nn'}(\Omega,\Omega')$ 
from $h^{K,\circ}_{nn'}(\Omega,\Omega')$, $n \neq n'$}
In Apps.~\ref{appI}-\ref{appIV}, we discuss the most general case
$K \subseteq S^{2}$. If $K = S^{2}$, the superscripts $K$ which
are used have to be dropped to be consistent with the rest of the
paper.

Using definiton~(\ref{eq53}), we have with Eq.~(\ref{eq54a})
\begin{align}
\label{eqA1}
h^{K}_{nn'}(\Omega,\Omega') = h^{K,\circ}_{nn'}(\Omega,\Omega')+
\frac{1}{|K|} \int\limits_{K} h^{K}_{nn'}(\Omega,\Omega') \, d\Omega 
+ \frac{1}{|K|} \int\limits_{K} h^{K}_{nn'}(\Omega,\Omega') \, d\Omega' -
\frac{1}{|K|^{2}} \iint\limits_{KK} h^{K}_{nn'}(\Omega,\Omega') \,
d\Omega \, d\Omega'
\end{align}
If we take $\iint_{K} \rho^{(1)}(\Omega)\, \ldots\,\rho^{(1)}(\Omega')
\,d\Omega \, d\Omega'$ on both sides of Eqs.~(\ref{eqA1}), it remains due to
Eqs.~(\ref{eq15})
\begin{align}
\label{eqA2}
\frac{1}{|K|^{2}} \iint\limits_{KK} h^{K}_{nn'}(\Omega,\Omega') \,
d\Omega \, d\Omega'  =
\iint\limits_{KK} \rho^{(1)}(\Omega) \,
 h^{K,\circ}_{nn'}(\Omega,\Omega') \,\rho^{(1)}(\Omega') \,d\Omega \, d\Omega'\,.
\end{align}
Only $\int_{K}  \ldots\,\rho^{(1)}(\Omega') \, d\Omega'$ applied to 
both sides of Eq.~(\ref{eqA1}) yields 
\begin{align}
\label{eqA3}
\frac{1}{|K|} \int\limits_{K} h^{K}_{nn'}(\Omega,\Omega') \, d\Omega' =
\frac{1}{|K|^{2}} \iint\limits_{KK} h^{K}_{nn'}(\Omega,\Omega')
d\Omega \, d\Omega'
- \int\limits_{K} h^{K,\circ}_{nn'}(\Omega,\Omega') \,\rho^{(1)}
(\Omega')\, d\Omega'
\end{align}
(analogously, $\int_{K} \rho^{(1)}(\Omega)\, \ldots\, d\Omega$
must be applied in the same manner). Then,~(\ref{eqA2}) can be used
with the first term on the rhs of~(\ref{eqA3}), 
and finally we find that
\begin{align}
\label{eqA4}
&h^{K}_{nn'}(\Omega,\Omega') = h^{K,\circ}_{nn'}(\Omega,\Omega')
-\int\limits_{K} h^{K,\circ}_{nn'}(\Omega,\Omega') \,\rho^{(1)}
(\Omega')\, d\Omega'
-\int\limits_{K} \rho^{(1)}
(\Omega)\, h^{K,\circ}_{nn'}(\Omega,\Omega') \, d\Omega
\nonumber\\
+&\iint\limits_{KK} \rho^{(1)}(\Omega) \,
 h^{K,\circ}_{nn'}(\Omega,\Omega') \,\rho^{(1)}(\Omega') \,d\Omega \,
d\Omega' = \iint\limits_{KK} \left(\delta\,(\Omega|\Omega'')-
 \rho^{(1)}(\Omega'')\right) h^{K,\circ}_{nn'}(\Omega'',\Omega''') 
\left( \delta\,(\Omega'''|\Omega') - \rho^{(1)}(\Omega''')\right) d\Omega''
\,d\Omega'''\,.
\end{align}
Therefore, $h^{K}_{nn'}(\Omega,\Omega')$ is determined by 
$h^{K,\circ}_{nn'}(\Omega,\Omega')$ via Eq.~(\ref{eqA4}). It is easy to prove
that this $h^{K}_{nn'}(\Omega,\Omega')$ is the only one leading
to the given $h^{K,\circ}_{nn'}(\Omega,\Omega')$ and
simultaneously fulfilling Eqs.~(\ref{eq15}),
i.e. $h^{\circ,K}_{nn'}(\Omega,\Omega')$ determines 
$h^{K}_{nn'}(\Omega,\Omega')$ uniquely.
\end{widetext}
\section{\label{appII}Calculation of the matrix elements
$g^{K}_{nn',\lambda\lambda'}$, $n \neq n'$}
For the PY approximation in matrix form~(\ref{eq50}),
the matrix elements $g^{K}_{nn',\lambda\lambda'} =
1^{K}_{\lambda\lambda'} + h^{K}_{nn',\lambda\lambda'}$ are needed. 
In the following, it is convenient to introduce the projection
operator $P^{K} = (P^{K})^{2} = (P^{K})^{\dagger}$, defined on
$S^2 \times S^2$ as
\begin{subequations}
\label{eqB1}
\begin{eqnarray}
\label{eqB1a}
P^{K}(\Omega,\Omega') & = &
\left\{
\begin{array}{c}
\delta (\Omega|\Omega')\,, \quad \Omega,\Omega' \in K, \\
0 \, , \text{     otherwise}
\end{array}
\right . \\
\label{eqB1b}
P^{K}_{\lambda\lambda'} & = &  i^{l'-l}\int\limits_{K} 
 Y_{\lambda}^{*}(\Omega) \, Y_{\lambda'}(\Omega) \, d\Omega \,,
\end{eqnarray}
\end{subequations}
and
\begin{subequations}
\label{eqB2}
\begin{align}
\label{eqB2a}
d(\Omega,\Omega') & = 4\pi\,
\rho^{(1)}(\Omega)\,\delta(\Omega|\Omega') \\
\label{eqB2b}
d_{\lambda\lambda'} & = 4\pi\, i^{l'-l} \int\limits_{K} 
\rho^{(1)}(\Omega)\,Y_{\lambda}^{*}(\Omega) \, Y_{\lambda'}(\Omega) \, d\Omega .
\end{align}
Using (\ref{eqB2}), the $\lambda$ transform of $D(\Omega,\Omega')$
(see Eq.~(\ref{eq18})) can be represented as
\begin{equation}
\label{eqB2c}
D_{\lambda\lambda'} = d_{\lambda\lambda'} -  d_{\lambda,00}
\,d_{00,\lambda'}
\end{equation}
\end{subequations}
Though defined on $S^2 \times S^2$, $d = d^{K}$ can be
assumed, since $d(\Omega,\Omega') = 0$ on 
$(S^2 \times S^2) \setminus (K \times K)$. The inverse function for 
$d$ on $K$ is $\frac{1}{4\pi}\frac{\delta(\Omega|\Omega')}{\rho^{(1)}(\Omega)}$.
Note that for $K = S^{2}$ one has $P^{S^{2}}_{\lambda\lambda'} =
\delta_{\lambda,00} \,\delta_{00,\lambda'}$.

Now we calculate $g^{K}_{nn',\lambda\lambda'}$. First, 
the matrix elements $1^{K}_{\lambda\lambda'}$ are given
\begin{align}
\label{eqB3}
1^{K}_{\lambda\lambda'} = i^{l'-l} \iint\limits_{KK} Y_{\lambda}^{*}(\Omega)
\, Y_{\lambda'} (\Omega') \, d \Omega \, d \Omega' = 4\pi\,
P^{K}_{\lambda,00}\, P^{K}_{00,\lambda'} \,.
\end{align}
Next, Eq.~(\ref{eqA4}) is used. The matrices ${\bf
h}^{K,\circ}_{nn'}$ are delivered by the OZ equation. According 
to~(\ref{eqB3}), the matrix belonging to $\iint_{K} \rho^{(1)}(\Omega) \,
 h^{K,\circ}_{nn'}(\Omega,\Omega') \,\rho^{(1)}(\Omega') \,d\Omega \,
d\Omega'$ is
\begin{align}
\label{eqB4}
&\left( \iint\limits_{KK} \rho^{(1)}(\Omega) \,
 h^{K,\circ}_{nn'}(\Omega,\Omega') \,\rho^{(1)}(\Omega') \,d\Omega \,
d\Omega' \right)_{\lambda\lambda'} \nonumber \\
&=  4\pi \,P^{K}_{\lambda,00}\, P^{K}_{00,\lambda'} 
\iint\limits_{KK} \rho^{(1)}(\Omega) \,
 h^{K,\circ}_{nn'}(\Omega,\Omega') \,\rho^{(1)}(\Omega') \,d\Omega \,
d\Omega' \,,
\end{align}
where by use of~(\ref{eqB2b}) one easily computes that (using the summation rule)
\begin{align}
\label{eqB5}
\iint\limits_{KK} \rho^{(1)}&(\Omega) \,
h^{K,\circ}_{nn'}(\Omega,\Omega') \,\rho^{(1)}(\Omega') \,d\Omega \,
d\Omega' \nonumber \\
&=\frac{1}{4\pi} \,
d_{00,\lambda} \, h^{K,\circ}_{nn',\lambda\lambda'} \, d_{\lambda',00}\,.
\end{align}
The matrix elements $(\int_{K} h^{K,\circ}_{nn'}
(\Omega,\Omega') \,\rho^{(1)} (\Omega')\, d\Omega')_{\lambda\lambda'}$ 
are also straightforward. Since the angular function depends not on
$\Omega'$, $ \sqrt{4\pi} \, P^{K}_{00,\lambda'}$ must occur, as in
Eq.~(\ref{eqB3}). The dependence on $\Omega$ and the integrals are
represented by $\sqrt{4\pi}^{-1} \, h^{K,\circ}_{nn',\lambda\lambda''} \,
d_{\lambda'',00}$, so that we finally have
\begin{align}
\label{eqB6}
\left(\int\limits_{K} h^{K,\circ}_{nn'}
(\Omega,\Omega') \,\rho^{(1)} (\Omega') \,
d\Omega'\right)_{\lambda\lambda'} = 
h^{K,\circ}_{nn',\lambda\lambda''} \,
d_{\lambda'',00} \, P^{K}_{00,\lambda'} \,.
\end{align}
Altogether, the matrix elements $h^{K}_{nn',\lambda\lambda'}$, which
must be completed by~(\ref{eqB3}) to give
$g^{K}_{nn',\lambda\lambda'}$, are
\begin{widetext}
\begin{align}
\label{eqB7}
h^{K}_{nn',\lambda\lambda'}&= h^{K,\circ}_{nn',\lambda\lambda'}
-\,h^{K,\circ}_{nn',\lambda\lambda''} \,
d_{\lambda'',00} \, P^{K}_{00,\lambda'}
-P^{K}_{\lambda,00} \,d_{00,\lambda''}
\,h^{K,\circ}_{nn',\lambda''\lambda'} 
+P^{K}_{\lambda,00} \,
d_{00,\lambda''} \, h^{K,\circ}_{nn',\lambda''\lambda'''} \,
d_{\lambda''',00}\, P^{K}_{00,\lambda'}
\nonumber\\
&= \left(P^{K}_{\lambda\lambda''} - P^{K}_{\lambda,00}
d_{00,\lambda''} \right)h^{K,\circ}_{nn',\lambda''\lambda'''}
\left(P^{K}_{\lambda'''\lambda'} - P^{K}_{00,\lambda'}
d_{\lambda''',00} \right)
\end{align}
\section{\label{appIII}Calculation of $f^{K}_{nn',\lambda\lambda'}$,
$n \neq n'$}
Contrary to the lattice correlation functions,
$f_{nn'}(\Omega,\Omega')$ is not affected by the lattice and
refers exclusively to what happens between two
particles. Therefore, it is advantageous to use the $r$-frame for the
calculation of the matrix elements, i.e. the coordinate system in
which the connecting line of the particles coincides with the
$z$-axis, and later to transform them back to the laboratory
system. For $K = S^{2}$ in the $r$-frame, the matrix elements have similar properties 
as in~\cite{SS1997}:
\begin{equation}
\label{eqC1}
f^{r}_{nn',lm,l'm'} = f^{r}_{nn',ll'm} \, \delta_{mm'}
\,,\quad\quad f^{r}_{nn',ll'm} =  (-1)^{l+l'} f^{r^{*}}_{nn',ll'm}
\,,\quad\quad f^{r}_{nn',ll'm} = f^{r}_{nn',ll'-m}\,.
\end{equation}

In the following, the abbreviation $Q_{lm}(\theta) = (-1)^{m}
\left(\frac{2l+1}{4\pi}\right)^{\frac{1}{2}}
\left(\frac{(l-m)!}{(l+m)!} \right)^{\frac{1}{2}} \sin\theta \,
P_{lm}(\cos\theta)$ is used, where we define $P_{l,-m}(\cos\theta)
= (-1)^{m} \frac{(l-m)!}{(l+m)!} \,P_{lm}(\cos\theta)$ 
to cover all possible values of $m$. Then it is
\begin{align}
\label{eqC2}
f^{r}_{nn',ll'm} 
= i^{l'-l} \int\limits_{0}^{2\pi} \int\limits_{0}^{2\pi}
\int\limits_{0}^{\pi} \int\limits_{0}^{\pi} 
f^{r}(|{\bf x}_{nn'}|,\theta_{r},\theta_{r}',\phi_{r}'-\phi_{r}) 
\, Q_{lm}(\theta_{r}) \,Q_{l'm}(\theta_{r}') \,
\text{e}^{im(\phi_{r}'-\phi_{r})} \, d\theta_{r} \, d\theta_{r}' \,d\phi_{r} \,d\phi_{r}'
\end{align}
Switching to the new variables $\phi_{1} = \phi_{r}'- \phi_{r}$ and 
$\phi_{2} = \phi_{r}'+\phi_{r}$ brings about a functional determinant
factor of $\frac{1}{2}$:
\begin{align}
\label{eqC3}
f^{r}_{nn',ll'm} = \frac{1}{2}\,i^{l'-l} \int\limits_{0}^{4\pi}
\int\limits_{-\Psi(\phi_{2})}^{\Psi(\phi_{2})}
\int\limits_{0}^{\pi} \int\limits_{0}^{\pi} 
f^{r}(|{\bf x}_{nn'}|,\theta_{r},\theta_{r}',\phi_{1}) 
\,Q_{lm}(\theta_{r}) \,Q_{l'm}(\theta_{r}') \,
\text{e}^{im \phi_{1}} \, d\theta_{r} \, d\theta_{r}' \,d\phi_{1} \,d\phi_{2}\,,
\end{align}
where $\Psi(\phi_{2}) = \phi_{2}$, $0 \leq \phi_{2} \leq 2\pi$, 
$\Psi(\phi_{2}) = 4\pi - \phi_{2}$, $2\pi \leq \phi_{2} \leq 4\pi$. 
Now, the symmetries $f^{r}(|{\bf x}_{nn'}|,\theta_{r},\theta_{r}',\phi_{1}) = 
f^{r}(|{\bf x}_{nn'}|,\theta_{r},\theta_{r}',-\phi_{1}) = 
f^{r}(|{\bf x}_{nn'}|,\theta_{r},\theta_{r}',2\pi-\phi_{1})$ can be used:
\begin{align}
\label{eqC4}
f^{r}_{nn',ll'm} & = i^{l'-l} \int\limits_{0}^{4\pi}
\int\limits_{0}^{\Psi(\phi_{2})}
\underbrace{\int\limits_{0}^{\pi} \int\limits_{0}^{\pi} 
f^{r}(|{\bf x}_{nn'}|,\theta_{r},\theta_{r}',\phi_{1}) 
\,Q_{lm}(\theta_{r}) \,Q_{l'm}(\theta_{r}') \,
\cos (m \phi_{1}) \, d\theta_{r} \, d\theta_{r}' \,}_{\circledast}d\phi_{1}
\,d\phi_{2} \nonumber \\
&= 2 \,i^{l'-l} \int\limits_{0}^{2\pi}
\int\limits_{0}^{\phi_{2}} \circledast \;d\phi_{1}
\,d\phi_{2} = 2 \,i^{l'-l} \int\limits_{0}^{2\pi}
\int\limits_{\phi_{1}}^{2\pi}
\circledast\;d\phi_{2}
\,d\phi_{1} = 2 \,i^{l'-l} \int\limits_{0}^{2\pi}
 (2\pi - \phi_{1}) \,
\circledast \;d\phi_{1} = 2\pi \,i^{l'-l} \int\limits_{0}^{2\pi}
\circledast \; d\phi_{1}  \nonumber \\
&= 4\pi \,i^{l'-l} \int\limits_{0}^{\pi}
\int\limits_{0}^{\pi} \int\limits_{0}^{\pi}
f^{r}(|{\bf x}_{nn'}|,\theta_{r},\theta_{r}',\phi_{1}) 
\,Q_{lm}(\theta_{r}) \,Q_{l'm}(\theta_{r}') \,
\cos (m \phi_{1}) \, d\theta_{r} \, d\theta_{r}'
\,d\phi_{1} \,.
\end{align}
For particles of inversion symmetry, only the ($ll'$ even) matrix
elements do not vanish, and by using 
$f^{r}(|{\bf x}_{nn'}|,\theta_{r},\theta_{r}',\phi_{1}) = 
f^{r}(|{\bf x}_{nn'}|,\theta_{r},\pi-\theta_{r}',\pi+\phi_{1}) = \ldots$
for that case we find that 
\begin{align}
\label{eqC5}
f^{r}_{nn',ll'm} = 16\pi \,i^{l'-l} \int\limits_{0}^{\pi}
\int\limits_{0}^{\frac{\pi}{2}} \int\limits_{0}^{\frac{\pi}{2}}
f^{r}(|{\bf x}_{nn'}|,\theta_{r},\theta_{r}',\phi_{1}) 
\,Q_{lm}(\theta_{r}) \,Q_{l'm}(\theta_{r}') \,
\cos (m \phi_{1}) \, d\theta_{r} \, d\theta_{r}'
\,d\phi_{1} \,.
\end{align}

Finally, having transformed back to the laboratory frame and the
matrix elements $f_{nn',\lambda\lambda'}$ in hand, they can be projected 
due to ${\bf f}^{K}_{nn'} = {\bf P}^{K} \,{\bf f}_{nn'} \, {\bf P}^{K}$.
\end{widetext}
\section{\label{appIV}Calculation of
$c^{K}_{nn'}(\Omega,\Omega')$, $n \neq n'$, for hard particles}
The appendices~\ref{appIV} and~\ref{appV} are given in terms of
angular functions only. The discussion in terms of matrices is
much more complicated.

For hard particles, the values of $c^{K}_{nn'}(\Omega,\Omega')$ 
in the areas of overlap remain undetermined by 
Eq.~(\ref{eq48}), and the question arises if these 
values are also uniquely determined by $c^{K,\circ}_{nn'}(\Omega,\Omega')$. That this
indeed is true will be proven below.

Most important for the following considerations is that no space angle 
$\Omega'_{0} \in K$ can be found for which the pair
$(n\Omega,n'\Omega'_{0})$ has overlap for all $\Omega \in K$. 
Otherwise, it would be $\Omega'_{0} \notin K$. Note that, contrary to 
the case of overlap, for appropriate size of the
particles and special values of $\Omega'_{0} \in K$, the pair 
$(n\Omega,n'\Omega'_{0})$ has no overlap for all $\Omega \in K$. 
Given a solution $c^{K}_{nn'}(\Omega,\Omega')$ for which 
$c^{K,\circ}_{nn'}(\Omega,\Omega') = (R^{K}\,c^{K}_{nn'}\,R^{K})
(\Omega,\Omega')$, it is $c^{K}_{nn'}(\Omega,\Omega') = 0$,
if $(n\Omega,n'\Omega')$ has no
overlap. Eqs.~(\ref{eq53}),~(\ref{eq54a}) 
and the symmetries of the correlation functions given in Sec.~\ref{secIIb}
show that
\begin{align}
\label{eqD1}
c^{K}_{nn'}(\Omega,\Omega') = c^{K,\circ}_{nn'}(\Omega,\Omega') +
\hat{c}^{K}_{nn'}(\Omega) + \hat{c}^{K}_{n'n}(\Omega') - \hat{c}^{K}_{nn'}\,.
\end{align}
Let us fix $\Omega' = \Omega'_{1} \in K$ and let $K(\Omega'_{1})
\subset K$ be the area such that $(n\Omega,n'\Omega'_{1})$ has no
overlap for all  $\Omega \in K(\Omega'_{1})$. Then it follows from
Eq.~(\ref{eqD1}) 
\begin{align}
\label{eqD2}
\hat{c}^{K}_{nn'}(\Omega) = - c^{K,\circ}_{nn'}(\Omega,\Omega'_{1}) -
\hat{c}^{K}_{n'n}(\Omega'_{1}) +\hat{c}^{K}_{nn'}\,.
\end{align}
for all $\Omega \in K(\Omega'_{1})$, because of
$c^{K}_{nn'}(\Omega,,\Omega'_{1}) =0$. Since
$c^{K,\circ}_{nn'}(\Omega,\Omega'_{1})$
is known, we have found $\hat{c}^{K}_{nn'}(\Omega)$ on
$K(\Omega'_{1})$ up to a constant depending on $\Omega'_{1}$. Next
we choose $\Omega' = \Omega'_{2} \in K$ and $\Omega'_{2} \neq
\Omega'_{1}$. Then $\hat{c}^{K}_{nn'}(\Omega)$ is obtained for
$\Omega \in K(\Omega'_{2})$ up to a constant depending on
$\Omega'_{2}$. This can be continued by choosing $\Omega'_{i}$, $i
=1,2,\ldots, M$ such that
$\cup_{i} \, K(\Omega'_{i}) =
K$. In that case one can construct $\hat{c}^{K}_{nn'}(\Omega)$ on
$K$, for example in the form $\hat{c}^{\circ,K}_{nn'}(\Omega)$ having
vanishing constant part. The same is done for
$\hat{c}^{\circ,K}_{n'n}(\Omega')$. Finally, we have
\begin{align}
\label{eqD3}
c^{K}_{nn'}(\Omega,\Omega') = c^{K,\circ}_{nn'}(\Omega,\Omega') +
\hat{c}^{K,\circ}_{nn'}(\Omega) + \hat{c}^{K,\circ}_{n'n}(\Omega') - 
\hat{\hat{c}}^{K}_{nn'}\,,
\end{align}
where $\hat{\hat{c}}^{K}_{nn'}$ follows immediately if~(\ref{eqD3}) 
is used with any $\Omega,\Omega'$ such that $(n\Omega,n'\Omega')$ 
has no overlap.
\section{\label{appV}Some remarks on how the case $K \subset S^{2}$
can be treated}
In this appendix, we want to answer two questions concerning the case
of hard core particles, whose geometry leads to $K \subset
S^{2}$. First, assuming finite interaction if the ``hard cores'' 
of such particles overlap, i.e. 
\begin{equation}
\label{eqE1}
V_{nn'}(\Omega,\Omega') = 
V < \infty\, \text{ (overlap)}\,,
\end{equation}
what happens to the solutions $h^{V}_{nn'}(\Omega,\Omega')$ and 
$c^{V}_{nn'}(\Omega,\Omega')$ of the OZ/PY 
equations~(\ref{eq46a}) and~(\ref{eq47}), together with Eqs.~(\ref{eqA4})
and~(\ref{eq32a}), which are 
defined on $S^{2} \times S^{2}$, if we take the limit 
$V \longrightarrow \infty\,$? Second, does a 
$S^{2} \times S^{2}$-solution $h^{\infty}_{nn'}(\Omega,\Omega')$ and  
$c^{\infty}_{nn'}(\Omega,\Omega')$ of the OZ/PY
equations for $V=\infty$ lead to the same static 
structure factors as the solution $h^{K}_{nn'}(\Omega,\Omega')$ and
$c^{K}_{nn'}(\Omega,\Omega')$ of Sec.~\ref{secIIIb}?

The first question can be answered using the fact that the function 
$D^{V}(\Omega,\Omega')$ (cf. Eq.~(\ref{eq18})) and the Mayer function Eq.~(\ref{eq48}) 
behave continuously if we take the limit $V \longrightarrow \infty$:
\begin{subequations}
\label{eqE2}
\begin{align}
\lim_{V \to \infty} D^{V}(\Omega,\Omega') &= D^{\infty}(\Omega,\Omega') \\
\lim_{V \to \infty} f_{nn'}^{V}(\Omega,\Omega') &= f_{nn'}^{\infty}(\Omega,\Omega')
\end{align}
\end{subequations}
This is true for $D^{V}(\Omega,\Omega')$ since the one-particle
density $\rho^{(1)}(\Omega)$ varies continuously with $V$.
If also all other functions exist in the limit $V = \infty$ and 
if the limit interchanges with integration, we can write down
Eqs.~(\ref{eq46a}),~(\ref{eq47}),~(\ref{eqA4}) 
and~(\ref{eq32a}) including the functions
$h^{V}_{nn'}(\Omega,\Omega')$, $h^{V,\circ}_{nn'}(\Omega,\Omega')$,
$c^{V}_{nn'}(\Omega,\Omega')$, $c^{V,\circ}_{nn'}(\Omega,\Omega')$,
$D^{V}(\Omega,\Omega')$ and $f^{V}_{nn'}(\Omega,\Omega')$ for $V <
\infty$ and take this limit. Then it is easy to see that the limiting 
direct and total correlation functions are a set of functions 
$h^{\infty}_{nn'}(\Omega,\Omega')$, $c^{\infty}_{nn'}(\Omega,\Omega')$, 
i.e. they solve the OZ/PY equations for $V = \infty$ on $S^{2} \times
S^{2}$. Note that, as $h^{V}_{nn'}(\Omega,\Omega')$,
$h^{\infty}_{nn'}(\Omega,\Omega')$ fulfills the
constraints given by Eqs.~(\ref{eq15}).

Now we investigate if a solution $h^{\infty}_{nn'}(\Omega,\Omega')$ and
$c^{\infty}_{nn'}(\Omega,\Omega')$ leads to the correct static structure factors
Eqs.~(\ref{eq40}),~(\ref{eq41}) for $V = \infty$. For this, since
$D^{\infty}(\Omega,\Omega') = (D^{\infty}\,R^{K})(\Omega,\Omega') =
(R^{K}\,D^{\infty})(\Omega,\Omega') = (R^{K}\,D^{\infty}\,R^{K})(\Omega,\Omega')
$, with $R^{K}(\Omega,\Omega')$ due to Eq.~(\ref{eq54}),
we have to prove that $(R^{K} \,h^{\infty,\circ}_{nn'}\,R^{K})(\Omega,\Omega') =
h^{K,\circ}_{nn'}(\Omega,\Omega')$, $\Omega,\Omega' \in K$, 
where the former symbol $\circ$ refers to removed constant parts with
respect to $S^{2}$ and the latter with respect to $K$. This can be
done as follows.

a) It is tempting to assume that the functions
$h^{\infty}_{nn'}(\Omega,\Omega')$ and $c^{\infty}_{nn'}(\Omega,\Omega')$ 
restricted to $K$ match the functions $h^{K}_{nn'}(\Omega,\Omega')$ and
$c^{K}_{nn'}(\Omega,\Omega')$ of Sec.~\ref{secIIIb}. Then, the PY 
approximation~(\ref{eq47}) restricted to $K$ is fulfilled trivially.
Additionaly, the restricted function $h^{\infty}_{nn'}(\Omega,\Omega')$
fulfills the constraints of Eqs.~(\ref{eq15}) (rewritten for $K$ instead
of $S^{2}$), since these integrals do not change 
if one restricts the integration domains to $K$
(because of $\rho^{(1)}(\Omega) = 0$ on $\bar{K}$). 

b) Because of
\begin{align}
\label{eqD10}
\int\limits_{K} & \left(\delta (\Omega|\Omega') - \frac{1}{K} \right)
\left\{ \int\limits_{S^2} \left(\delta (\Omega'|\Omega'') -
\frac{1}{4\pi} \right) f(\Omega'') \, d\Omega''
\right\} d\Omega' \nonumber \\
& = \int\limits_{K} \left(\delta (\Omega|\Omega') -
\frac{1}{K} \right) f(\Omega') \, d\Omega'
\end{align}
is the removal of the constant parts of a function with respect to
$S^{2}$ and a subsequent restriction to $K$ and removal of the
constant parts with respect to $K$ equivalent to restricting the
function to $K$ first and then to remove the constant parts
with respect to $K$.

c) Next we notice that the OZ equation~(\ref{eq46a}) for $S^{2}
\times S^{2}$ decouples for $V =\infty$ on $K \times K$ completely
from $\bar{K}$, as does the OZ equation of Sec.~\ref{secIIIb}. 

d) If we write down this decoupled OZ equation with its 
solution $h^{\infty,\circ}_{nn'}(\Omega,\Omega')$ and
$c^{\infty,\circ}_{nn'}(\Omega,\Omega')$, restrict it to $K \times K$ 
and apply the projection operator $R^{K}(\Omega,\Omega')$
from both sides, it involves the functions $(R^{K}
\,h^{\infty,\circ}_{nn'}\,R^{K})(\Omega,\Omega')$ and $(R^{K}
\,c^{\infty,\circ}_{nn'}\,R^{K})(\Omega,\Omega')$. Due to b), this
is the OZ equation on $K \times K$, written with the functions 
$h^{\infty}_{nn'}(\Omega,\Omega')$ and
$c^{\infty}_{nn'}(\Omega,\Omega')$, which are restricted to $K$ and have
removed constant parts with respect to $K$.

Now we have shown that the initial assumption made in a) was right.
Therefore, to solve the OZ/PY equations for the case $K
\subset S^{2}$, one can solve these equations also on $S^{2}$ for $V =
\infty$, since the irrelevant $\bar{K}$-parts of the involved
functions are projected out. We have also made plausible that the limiting
process from finite to hard core interaction leads to such a solution
on $S^{2}$, and that therefore the structure factors behave
continuously for $V \longrightarrow \infty$. This is exactly what one
expects by intuition.


\begin{thebibliography}{20}
\expandafter\ifx\csname natexlab\endcsname\relax\def\natexlab#1{#1}\fi
\expandafter\ifx\csname bibnamefont\endcsname\relax
  \def\bibnamefont#1{#1}\fi
\expandafter\ifx\csname bibfnamefont\endcsname\relax
  \def\bibfnamefont#1{#1}\fi
\expandafter\ifx\csname citenamefont\endcsname\relax
  \def\citenamefont#1{#1}\fi
\expandafter\ifx\csname url\endcsname\relax
  \def\url#1{\texttt{#1}}\fi
\expandafter\ifx\csname urlprefix\endcsname\relax\def\urlprefix{URL }\fi
\providecommand{\bibinfo}[2]{#2}
\providecommand{\eprint}[2][]{\url{#2}}

\bibitem[{\citenamefont{{J.-P. Hansen, I. R. McDonald}}(1990)}]{HMD1990}
\bibinfo{author}{\bibnamefont{{J.-P. Hansen, I. R. McDonald}}},
  \emph{\bibinfo{title}{Theory of {S}imple {L}iquids}}
  (\bibinfo{publisher}{Academic Press}, \bibinfo{address}{San Diego},
  \bibinfo{year}{1990}), \bibinfo{edition}{reprint 2nd} ed.

\bibitem[{\citenamefont{{C. G. Gray, K. E. Gubbins}}(1984)}]{GG1984}
\bibinfo{author}{\bibnamefont{{C. G. Gray, K. E. Gubbins}}},
  \emph{\bibinfo{title}{Theory of {M}olecular {F}luids}},
  vol.~\bibinfo{volume}{1} (\bibinfo{publisher}{Clarendon Press},
  \bibinfo{address}{Oxford}, \bibinfo{year}{1984}).

\bibitem[{\citenamefont{{J. Ram, R. C. Singh, Y. Singh}}(1994)}]{RSS1994}
\bibinfo{author}{\bibnamefont{{J. Ram, R. C. Singh, Y. Singh}}},
  \bibinfo{journal}{Phys. Rev. E} \textbf{\bibinfo{volume}{49}},
  \bibinfo{pages}{5117} (\bibinfo{year}{1994}).

\bibitem[{\citenamefont{{M. Letz, A. Latz}}(1999)}]{LL1999}
\bibinfo{author}{\bibnamefont{{M. Letz, A. Latz}}}, \bibinfo{journal}{Phys.
  Rev. E} \textbf{\bibinfo{volume}{60}}, \bibinfo{pages}{5865}
  (\bibinfo{year}{1999}).

\bibitem[{\citenamefont{{J. D. Wright}}(2001)}]{JDW2001}
\bibinfo{author}{\bibnamefont{{J. D. Wright}}}, \emph{\bibinfo{title}{Molecular
  {C}rystals}} (\bibinfo{publisher}{Cambrigde University Press},
  \bibinfo{address}{Cambridge}, \bibinfo{year}{2001}).

\bibitem[{\citenamefont{{J. N. Sherwood}}(1979)}]{SHER1979}
\bibinfo{editor}{\bibnamefont{{J. N. Sherwood}}}, ed.,
  \emph{\bibinfo{title}{The {P}lastically {C}rystalline {S}tate}}
  (\bibinfo{publisher}{John {W}iley \& {S}ons}, \bibinfo{address}{Chichester},
  \bibinfo{year}{1979}).

\bibitem[{\citenamefont{{M. H. M\"user}}(1998)}]{MUE1998}
\bibinfo{author}{\bibnamefont{{M. H. M\"user}}},
  \bibinfo{journal}{Ferroelectrics} \textbf{\bibinfo{volume}{208}},
  \bibinfo{pages}{293} (\bibinfo{year}{1998}).

\bibitem[{\citenamefont{{R. M. Lynden-Bell, K. H. Michel}}(1994)}]{LM1994}
\bibinfo{author}{\bibnamefont{{R. M. Lynden-Bell, K. H. Michel}}},
  \bibinfo{journal}{Rev. Mod. Phys.} \textbf{\bibinfo{volume}{66}},
  \bibinfo{pages}{721} (\bibinfo{year}{1994}).

\bibitem[{\citenamefont{{M. Yvinec, R. Pick}}(1980)}]{YP1980}
\bibinfo{author}{\bibnamefont{{M. Yvinec, R. Pick}}}, \bibinfo{journal}{J.
  Physique} \textbf{\bibinfo{volume}{41}}, \bibinfo{pages}{1045, 1053}
  (\bibinfo{year}{1980}).

\bibitem[{\citenamefont{{W. Breymann, R. Pick}}(1988)}]{BP1988}
\bibinfo{author}{\bibnamefont{{W. Breymann, R. Pick}}},
  \bibinfo{journal}{Europhys. Lett.} \textbf{\bibinfo{volume}{6}},
  \bibinfo{pages}{227} (\bibinfo{year}{1988}).

\bibitem[{\citenamefont{{W. Breymann, R. Pick}}(1989)}]{BP1989}
\bibinfo{author}{\bibnamefont{{W. Breymann, R. Pick}}}, \bibinfo{journal}{J.
  Chem. Phys.} \textbf{\bibinfo{volume}{91}}, \bibinfo{pages}{3119}
  (\bibinfo{year}{1989}).

\bibitem[{\citenamefont{{W. Breymann, R. Pick}}(1994)}]{BP1994}
\bibinfo{author}{\bibnamefont{{W. Breymann, R. Pick}}}, \bibinfo{journal}{J.
  Chem. Phys.} \textbf{\bibinfo{volume}{100}}, \bibinfo{pages}{2232}
  (\bibinfo{year}{1994}).

\bibitem[{\citenamefont{{R. Schilling, T. Scheidsteger}}(1997)}]{SS1997}
\bibinfo{author}{\bibnamefont{{R. Schilling, T. Scheidsteger}}},
  \bibinfo{journal}{Phys. Rev. E} \textbf{\bibinfo{volume}{56}},
  \bibinfo{pages}{2932} (\bibinfo{year}{1997}).

\bibitem[{\citenamefont{{T. Franosch, M. Fuchs, W. G\"otze, M. R. Mayr, A. P.
  Singh}}(1997)}]{FFG1997}
\bibinfo{author}{\bibnamefont{{T. Franosch, M. Fuchs, W. G\"otze, M. R. Mayr,
  A. P. Singh}}}, \bibinfo{journal}{Phys. Rev. E}
  \textbf{\bibinfo{volume}{56}}, \bibinfo{pages}{5659} (\bibinfo{year}{1997}).

\bibitem[{\citenamefont{{F. C. von der Lage, H. A. Bethe}}(1947)}]{LB1947}
\bibinfo{author}{\bibnamefont{{F. C. von der Lage, H. A. Bethe}}},
  \bibinfo{journal}{Phys. Rev.} \textbf{\bibinfo{volume}{71}},
  \bibinfo{pages}{612} (\bibinfo{year}{1947}).

\bibitem[{\citenamefont{{J. Vieillard-Baron}}(1972)}]{VB1972}
\bibinfo{author}{\bibnamefont{{J. Vieillard-Baron}}}, \bibinfo{journal}{J.
  Chem. Phys.} \textbf{\bibinfo{volume}{56}}, \bibinfo{pages}{4729}
  (\bibinfo{year}{1972}).

\bibitem[{\citenamefont{{A. Saupe}}(1964)}]{ST1964}
\bibinfo{author}{\bibnamefont{{A. Saupe}}}, \bibinfo{journal}{Z. Naturforsch.
  A} \textbf{\bibinfo{volume}{19A}}, \bibinfo{pages}{161}
  (\bibinfo{year}{1964}).

\bibitem[{\citenamefont{{C. Renner, H. L\"owen, J. L. Barrat}}(1995)}]{RLB1995}
\bibinfo{author}{\bibnamefont{{C. Renner, H. L\"owen, J. L. Barrat}}},
  \bibinfo{journal}{Phys. Rev. E} \textbf{\bibinfo{volume}{52}},
  \bibinfo{pages}{5091} (\bibinfo{year}{1995}).

\bibitem[{\citenamefont{{D. Frenkel, B. M. Mulder, J. P. Mc
  Tague}}(1984)}]{FMT1984}
\bibinfo{author}{\bibnamefont{{D. Frenkel, B. M. Mulder, J. P. Mc Tague}}},
  \bibinfo{journal}{Phys. Rev. Lett.} \textbf{\bibinfo{volume}{52}},
  \bibinfo{pages}{287} (\bibinfo{year}{1984}).

\bibitem[{\citenamefont{{J. K. Percus}}(1962)}]{P1962}
\bibinfo{author}{\bibnamefont{{J. K. Percus}}}, \bibinfo{journal}{Phys. Rev.
  Lett.} \textbf{\bibinfo{volume}{8}}, \bibinfo{pages}{462}
  (\bibinfo{year}{1962}).

\end{thebibliography}

\end{document}